\documentclass[journals,article,accept,moreauthors,pdftex,electronics]{Definitions/mdpi}

\firstpage{1} 
\pubvolume{10}
\issuenum{1349}
\articlenumber{1349}
\pubyear{2021}
\copyrightyear{2020}
\datereceived{30 March 2021} 
\dateaccepted{1 June 2021} 
\datepublished{5 June 2021} 
\hreflink{https://\linebreak doi.org/10.3390/electronics10111349} 

\Title{Stochastic Restoration of Heavily Compressed Musical Audio using Generative Adversarial Networks}

\TitleCitation{Stochastic Restoration of Heavily Compressed Musical Audio using Generative Adversarial Networks}

\Author{Stefan Lattner $^{1}$ and Javier Nistal $^{1}$}

\AuthorNames{Stefan Lattner, Javier Nistal}

\AuthorCitation{Lattner, S.; Nistal, J.}

\address{%
$^{1}$ \quad Sony Computer Science Laboratories (CSL), 75005 Paris, France; javier.nistalhurle@sony.com}

\corres{Correspondence: stefan.lattner@sony.com or me@stefanlattner.at}

\abstract{Lossy audio codecs compress (and decompress) digital audio streams by removing information that tends to be inaudible in human perception. Under high compression rates, such codecs may introduce a variety of impairments in the audio signal. Many works have tackled the problem of audio enhancement and compression artifact removal using deep learning techniques. However, only a few works tackle the restoration of \emph{heavily compressed} audio signals in the \emph{musical domain}. In such a scenario, there is no unique solution for the restoration of the original signal. 
Therefore, in this study, we test a \emph{stochastic} generator of a Generative Adversarial Network (GAN) architecture for this task. Such a stochastic generator, conditioned on highly compressed musical audio signals, could one day generate outputs indistinguishable from high-quality releases. Therefore, the present study may yield insights into more efficient musical data storage and transmission. We train stochastic and deterministic generators on MP3-compressed audio signals with 16, 32, and 64 kbit/s. We perform an extensive evaluation of the different experiments utilizing objective metrics and listening tests. We find that the models can improve the quality of the audio signals over the MP3 versions for 16 and 32 kbit/s and that the stochastic generators are capable of generating outputs that are closer to the original signals than those of the deterministic generators.}


\usepackage[utf8]{inputenc}
\usepackage{amsmath}
\usepackage{amssymb}
\usepackage{graphicx}
\usepackage{rotating}
\usepackage{url}
\usepackage{booktabs}
\DeclareMathOperator{\sign}{sign}

\begin{document}

\maketitle

\begin{paracol}{2}
\switchcolumn
\vspace{-80pt}
\section{Introduction}
The introduction of MP3 (i.e., MPEG-1 layer 3 \cite{mpeg-1}) was transformative in how music was stored, transmitted, and shared in digital devices and on the internet. MP3 players, sharing platforms and streaming resulted directly from the possibility to considerably compress audio data without noticeable perceptual compromises. Compared to \emph{lossless} audio coding formats, which allow for a perfect reconstruction of the original PCM audio signal, \emph{lossy} formats (like MP3) typically lead to better compression by ignoring the parts of the signal to which humans are less sensitive. This process is also called \emph{perceptual coding} which takes into account the physio- and psychological abilities of the human auditory perception, resulting in so-called psychoacoustic models \cite{Karlheinz}.


While there exist several different lossy audio codecs (e.g., AAC, Opus, Vorbis, AMR), MP3 is undoubtedly the most commonly used. It is built upon an analysis filter bank and a subsequent computation of the modified discrete cosine transform (MDCT). In parallel, the signal is analyzed based on a perceptual model that exploits the psychoacoustic phenomena of \emph{auditory masking} to determine sound events in the audio signal that are considered to be beyond human hearing capabilities. Based on this information, the spectral components are quantized with specific resolution and coded with variable bit allocation while keeping the noise introduced in this process below the masking thresholds \cite{DBLP:journals/tce/Musmann06}. This process may introduce a variety of deficiencies when configured with incorrect or very extreme parameters. For example, under large compression rates, high-frequency content is susceptible to being removed, resulting in a  \emph{bandwidth loss}. \emph{Pre-echoes} can occur when decoding very sudden sound events for which the quantization noise spreads out over the synthesis window and consequently precede the event causing the noise. Other common artifacts are so-called \emph{swirlies} \cite{sos}, characterized by fast energy fluctuations in the low-level frequency content of the sound. Furthermore, there are other problems related to MP3 compression such as \emph{double-speak} as well as a general loss of transient definition, transparency, loss of detail clarity, and more \cite{sos}.

Many works exist which tackle the problem of audio enhancement, including the removal of compression artifacts. The most common recent methods used for these types of problems are based on deep learning. Typically, they focus on specific types of impairments present in the audio signals (e.g., reverberation \cite{DBLP:conf/icassp/WilliamsonW17}, bandwidth loss \cite{DBLP:journals/corr/abs-2010-11362}, or audio codec artifacts \cite{DBLP:journals/taslp/ZhaoLF19, Fisher2016WaveMedicCN, DBLP:conf/interspeech/SkoglundV20, DBLP:conf/icassp/BiswasJ20, porov2018music}). Also, different types of neural network architectures have been studied for these tasks. For example, Convolutional Neural Networks (CNNs) \cite{DBLP:conf/interspeech/ParkL17}, WaveNet-like architectures \cite{Fisher2016WaveMedicCN, DBLP:conf/waspaa/GuptaSAW19}, and UNets \cite{DBLP:conf/interspeech/IsikGPVHK20, DBLP:conf/interspeech/HuLLXZFWZX20}.
However, most of the works in this line of research tackle the enhancement of \emph{speech} signals \cite{DBLP:journals/taslp/ZhaoLF19, DBLP:conf/interspeech/SkoglundV20, DBLP:conf/waspaa/GuptaSAW19, DBLP:conf/icassp/BiswasJ20, Fisher2016WaveMedicCN, DBLP:conf/interspeech/ParkL17, DBLP:journals/taslp/KontioLA07, DBLP:conf/interspeech/IsikGPVHK20, DBLP:conf/interspeech/HuLLXZFWZX20, DBLP:conf/icassp/LiL15, DBLP:journals/taslp/XuDDL15}, and only a few publications exist for \emph{musical} audio restoration \cite{DBLP:conf/icassp/LagrangeG20, Miron2018HIGHFM, porov2018music, DBLP:journals/nca/DengSESZFO20}.
This focus on speech is understandable, given the wide range of speech enhancement techniques in telephony, automatic speech recognition, and hearing aids.
Also, compared to musical audio signals, speech signals are easier to study, as they are more homogeneous, narrow-band, and usually monophonic.
In contrast, musical audio signals, particularly in the popular music genre, are highly varied. It typically consists of multiple, superimposed sources, which can be of any type, including (polyphonic) tonal instruments, percussion, (singing) voice, and various sound effects. In addition, music is typically broad-band, containing frequencies spanning over the entire human hearing range.

Given that studies on deep learning-driven audio codec artifact removal for musical audio data are underrepresented in audio enhancement research, in this work, we attempt to provide some more insights into this task. We investigate the limits of a generative neural network model when dealing with a general popular music corpus comprising music released in the last seven decades. In particular, we are interested in the ability of the model to regenerate lost information of heavily compressed musical audio signals using a \emph{stochastic generator} (which is not very common in audio enhancement, with \cite{DBLP:conf/icassp/BiswasJ20, DBLP:conf/waspaa/MaitiM19} being some exceptions). This work is not only relevant for the restoration of MP3 data in existing (older) music collections. In the light of current developments in musical audio generation, where full songs can already be generated from scratch \cite{DBLP:journals/corr/abs-2005-00341_long}, musical audio enhancement may soon possess a much more \emph{generative} aspect. It has already been shown that strong generative models can enhance heavily corrupted speech through resynthesis with neural vocoders \cite{DBLP:conf/waspaa/MaitiM19}. Along these lines, examining a \emph{generative} (i.e., stochastic) decoder for heavily compressed audio signals may contribute to insights about more efficient musical data storage and transmission. Today, music streaming is increasingly common, which poses issues regarding energy consumption and consequently environmental sustainability. When accepting deviations from the original recording, higher compression rates could be reached with a generative decoder without perceptual compromises in the listening experience. Moreover, for heavily compressed audio signals, there is no single best solution to recover the original version. Therefore, it may be interesting for users to generate multiple recoveries and pick the one they like most.

We introduce a Generative Adversarial Network (GAN) \cite{goodfellow2014generative} architecture for the restoration of MP3-encoded musical audio signals. We train different stochastic and deterministic generators on MP3s with different compression rates. Using these models, we investigate if 1) restorations of the models considerably improve the MP3 versions, 2) if we can systematically pick samples among the outputs of the stochastic generators which are closer to the original than such of the deterministic generators, and 3) if the stochastic generators generally output higher-quality restorations than the deterministic generators. To that end, we perform an extensive evaluation of the different experiment setups utilizing objective metrics and listening tests. We find that the models are successful in points 1 and 2, but the \emph{random} outputs of the stochastic generators are approximately on a par (i.e., do not improve) the overall quality compared to the deterministic models (point 3).

The proposed GAN architecture is based on dilated convolutions with skip connections, combined with a novel concept which we call Frequency Aggregation Filters. These are convolutional filters spanning the whole frequency range, which contribute to the stability of the training and constitute a consequent take on the problem of non-local correlations in the frequency spectrum (see Section \ref{sec:freq_aggregation}). We also find that using so-called self-gating considerably reduces the memory requirement of the architecture by halving the number of input maps to each layer without degradation of the results (see Section \ref{sec:gating}). In order to prevent mode collapse, we propose a regularization that enforces a correlation between differences in the noise input and differences in the model output (see Section \ref{sec:mode_collapse}). As opposed to most other works (but in line with few other approaches using GANs \cite{nistal2020} and U-Net-based architectures \cite{DBLP:conf/interspeech/IsikGPVHK20, DBLP:conf/interspeech/HuLLXZFWZX20}), we input (and output) directly the (non-linearly scaled) complex-valued spectrum to the generator, eliminating the need to deal with phase information separately.

The rest of this paper is organised as follows. In Section \ref{sec:related_work} we revise previous works in bandwidth extension and audio enhancement. In Section \ref{sec:method} we describe in depth the proposed GAN architecture (Section \ref{sec:architecture}), the training procedure (Section \ref{sec:training_procedure}), the dataset (Section \ref{sec:data}) and the evaluation methods (Section \ref{sec:evaluation}). Finally, in Section \ref{sec:results} we present and discuss the results and conclude with suggestions for future work in Section \ref{sec:conclusion}.
Audio examples of the work are provided in the accompanying website \footnote{\url{https://sonycslparis.github.io/restoration_mdpi_suppl_mat/}}.

\section{Related Work}
\label{sec:related_work}
In this work, Generative Adversarial Networks (GANs) are employed to restore MP3-compressed musical audio signals to their original high-quality versions. This task falls into the intersection of audio enhancement and bandwidth extension. Therefore, we review works on both these domains.

\subsection{Bandwidth Extension}

Low-resolution audio data (i.e., audio signals with a sample rate lower than 44.1kHz) is generally preferable for storage or transmission over band-limited channels, like streaming music over the internet. Also, lossy audio encoders can significantly reduce the amount of information by removing high-frequency content, but at the expense of potentially hampering the perceived audio quality. In order to restore the quality of such truncated audio signals, bandwidth extension (BWE) methods aim to reconstruct the missing high-frequency content of an audio signal given its low-frequency content as input \cite{ABWE}. BWE is alternatively referred to as \emph{audio re-sampling} or \emph{sample-rate conversion} in the field of Digital Signal Processing (DSP), or as \emph{audio super-resolution} in the Machine Learning (ML) literature. 
Methods for BWE have been extensively studied in areas like audio streaming and restoration, mainly for legacy speech telephony communication systems \cite{DBLP:conf/interspeech/BansalRS05, DBLP:conf/waspaa/GuptaSAW19, DBLP:journals/taslp/KontioLA07, DBLP:conf/icassp/LiL15} or, less commonly, for degraded musical material \cite{DBLP:conf/icassp/LagrangeG20, Miron2018HIGHFM}. 

Pioneering works to speech BWE were originally algorithmic and operated based on a source-filter model. There, the problem of regenerating a wide-band signal is divided into finding an upper-band source and the corresponding spectral envelope, or filter, for that upper band. While methods for source generation were based on simple modulation techniques such as spectral folding and translation of a so-called low-resolution baseband \cite{DBLP:conf/icassp/MakhoulB79}, the efforts focused on estimating the filter or spectral envelope \cite{dietz2002spectral}. These works introduced the so-called spectral band replication (SBR) method, where the lower frequencies of the magnitude spectra are duplicated, transposed, and adjusted to fit the high-frequency content. Because in most use-cases for speech BWE the full transmission stack is controlled, most of these algorithmic methods rely on side information about the spectral envelope, obtained at the encoder from the full wide-band signal, and then transmitted within the bitstream for subsequent reconstruction at the decoder. 

Learning-based approaches to speech BWE rely on large models to learn dependencies across the lower and higher end of the frequency spectrum. Methods based on Non-negative Matrix Factorization (NMF) treat the spectrogram as a fixed set of non-negative bases learned from wide-band signals \cite{DBLP:conf/interspeech/BansalRS05}. These bases are fixed at test time and used to estimate the activation coefficients that best explain the narrow-band signal. The wide-band signal is then reconstructed by a linear combination of the base vectors weighted by the activations. These methods up-sample speech audio signals up to 22.05kHz efficiently but are sensitive to non-linear distortions due to the linear-mixing assumption. Dictionary-based methods can significantly improve the speech quality over the NMF approach by reconstructing the high-resolution audio signals as a non-linear combination of units from a pre-defined clean dictionary \cite{DBLP:conf/waspaa/MandelC15}, or by casting the problem as an l1-optimization of an analysis dictionary learned from wide-band data \cite{DBLP:conf/icdsp/DongWC15}. 

Early works on speech BWE using neural networks inherited the source-filter methodology found in previous works. By employing spectral folding to regenerate the wide-band signal, a simple NN is used to adjust the spectral envelope of the generated upper-band \cite{DBLP:journals/taslp/KontioLA07}. Direct estimation of the missing high-frequency spectrum was not extensively studied until the introduction of deeper architectures \cite{DBLP:conf/icassp/LiL15}. Advances in computer vision \cite{DBLP:journals/pami/DongLHT16, DBLP:conf/cvpr/IsolaZZE17} inspired the usage of highly expressive models to audio BWE, leading to significant improvements in the up-sampling ratio and quality of the reconstructed audio signal. Different approaches followed: by generating the missing time-domain samples in a process analogous to image super-resolution \cite{DBLP:conf/iclr/KuleshovEE17}, by inpainting the missing content in a time-frequency representation \cite{Miron2018HIGHFM}, or by combining information from both domains, preserving the phase information \cite{DBLP:conf/icassp/LimYXDH18}. Powerful auto-regressive methods for raw audio signals based on SampleRNN \cite{DBLP:journals/taslp/LingAGD18}, or WaveNet \cite{DBLP:conf/waspaa/GuptaSAW19} are able to increase the maximum resolution to 16 kHz and 24 kHz sample-rate, respectively, without neglecting phase information, as it is the case in most works operating in the frequency domain \cite{Miron2018HIGHFM, DBLP:conf/icassp/LagrangeG20, DBLP:conf/interspeech/BansalRS05, DBLP:conf/icassp/LiL15, DBLP:journals/corr/abs-2010-11362}. Most recent techniques using sophisticated transformer-based GANs can up-sample speech to full resolution audio at 44.1 kHz sample-rate \cite{DBLP:journals/corr/abs-2010-11362}.

\subsection{Audio Enhancement}
Audio signals may suffer from a wide variety of environmental adversities: e.g., sound recordings using low-fidelity devices or in noisy and reverberant spaces; degraded speech in mobile or legacy telephone communications systems; musical material from old recordings, or heavily compressed audio signals for streaming services. Audio enhancement aims to improve the quality of corrupted audio signals by removing noisy additive components and restoring distorted or missing content to recover the original audio signal. The field was first introduced for applications in noisy communication systems to improve the quality and intelligibility of speech signals \cite{speech_enhancement}. Many studies have been carried out on speech audio enhancement, e.g., for speech recognition, speaker identification and verification \cite{DBLP:conf/interspeech/Ortega-GarciaG96, DBLP:conf/icassp/SeltzerYW13, DBLP:conf/slt/KolboekTJ16}, hearing assistance devices \cite{cochlear, Jitong}, de-reverberation \cite{DBLP:conf/icassp/WilliamsonW17}, and so on. In the specific case of audio codec restoration, many different techniques exist for improvement of speech signals \cite{DBLP:journals/taslp/ZhaoLF19, Fisher2016WaveMedicCN, DBLP:conf/interspeech/SkoglundV20, DBLP:conf/icassp/BiswasJ20}, yet only few works attempt the restoration of heavily compressed \emph{musical} audio signals \cite{porov2018music, DBLP:journals/nca/DengSESZFO20}.

Classic speech enhancement methods follow multiple approaches, primarily based on analysis, modification, and synthesis of the noisy signal's magnitude spectrum and often omitting phase information. Popular strategies are categorized into spectral subtraction methods \cite{1163209}, Wiener-type filtering \cite{1163086}, statistical model-based \cite{168664} and subspace methods \cite{DBLP:journals/speech/DendrinosBC91}. These approaches have proven successful when the additive noise is stationary. However, under highly non-stationary noise or reduced signal-to-noise ratios (SNR), they introduce artificial residual noise. 

Recent deep learning approaches to speech enhancement outperform previous methods in terms of perceived audio quality, effectively reducing both stationary and non-stationary noise components. Popular methods learn non-linear mapping functions of noisy-to-clean spectrogram signals \cite{DBLP:journals/taslp/XuDDL15} or learn masks in a time-frequency domain representation \cite{DBLP:conf/icassp/WilliamsonW17, DBLP:conf/interspeech/IsikGPVHK20, DBLP:journals/taslp/WilliamsonWW16}. Many architectures have been proposed: basic feed-forward DNNs \cite{DBLP:journals/taslp/XuDDL15}, CNN-based \cite{DBLP:conf/interspeech/ParkL17}, RNN-based \cite{DBLP:conf/icassp/ErdoganHWR15}, and more sophisticated architectures based on WaveNet \cite{Fisher2016WaveMedicCN} or U-Net \cite{DBLP:conf/interspeech/IsikGPVHK20}. GANs are also increasingly popular in speech enhancement \cite{DBLP:conf/interspeech/PascualBS17, DBLP:conf/interspeech/PascualSB19, DBLP:journals/cssp/LiDSM18, DBLP:conf/icassp/DonahueLP18}. Pioneering works using GANs operated either on the waveform domain \cite{DBLP:conf/interspeech/PascualBS17} or on the magnitude STFT \cite{DBLP:conf/interspeech/MichelsantiT17}. Subsequent works mainly focused on the latter representation due to the reduced complexity compared to time-domain audio signals \cite{DBLP:journals/cssp/LiDSM18, DBLP:conf/icassp/DonahueLP18, DBLP:conf/icml/FuLTL19}. Recent works operating directly on the raw waveform were able to consider a broader type of signal distortions \cite{DBLP:conf/interspeech/PascualSB19} and to improve the reduction of artifacts over previous works \cite{DBLP:journals/spl/PhanMPCKVM20}. Successive  efforts were made to further reduce artefacts by, for example, taking into consideration human perception. Some works directly optimize over differentiable approximations of objective metrics such as PESQ \cite{DBLP:conf/icml/FuLTL19}. 
However, these metrics correlate poorly with human perception, and some works defined the objective metric in embedding spaces from related tasks \cite{DBLP:conf/interspeech/GermainCK19} or by matching deep features of real and fake batches in the critic's embedding space \cite{DBLP:conf/interspeech/SuJF20}.

The vast majority of the speech audio enhancement approaches mentioned above operate on the magnitude spectrum and ignore the phase information \cite{DBLP:journals/cssp/LiDSM18, DBLP:journals/nca/DengSESZFO20, Miron2018HIGHFM, DBLP:conf/icassp/DonahueLP18}. At synthesis, researchers often reuse the phase spectrum from the noisy signal, introducing audible artifacts that would be particularly annoying in musical audio signals. To address this, phase-aware models for speech enhancement use a complex ratio mask \cite{DBLP:journals/taslp/WilliamsonWW16}, or, as we have seen, operate directly in the waveform domain \cite{DBLP:conf/interspeech/PascualSB19, DBLP:journals/spl/PhanMPCKVM20}. Inspired by a recent work demonstrating that DNNs implementing complex operators \cite{DBLP:conf/iclr/TrabelsiBZSSSMR18} may outperform previous architectures in many audio-related tasks, new state-of-the-art performances were achieved on speech enhancement using complex representations of audio data \cite{DBLP:conf/interspeech/IsikGPVHK20, DBLP:conf/interspeech/HuLLXZFWZX20}. Recent work was able to further improve these approaches by introducing a complex convolutional block attention module (CCBAM) and a mixed loss function \cite{DBLP:journals/corr/abs-2102-01993}.

\section{Materials and Methods}
\label{sec:method}
In the following, we describe the experiment setup. This includes the model architecture (see Section \ref{sec:architecture}) and the training procedure (see Section \ref{sec:training_procedure}). Furthermore, the data used and the data representation are presented in Section \ref{sec:data}, and the objective and subjective evaluation methods are discussed in Section \ref{sec:evaluation}.

\subsection{Model Architecture}\label{sec:architecture}

\end{paracol}
\begin{figure}
\widefigure
\includegraphics[width=\textwidth]{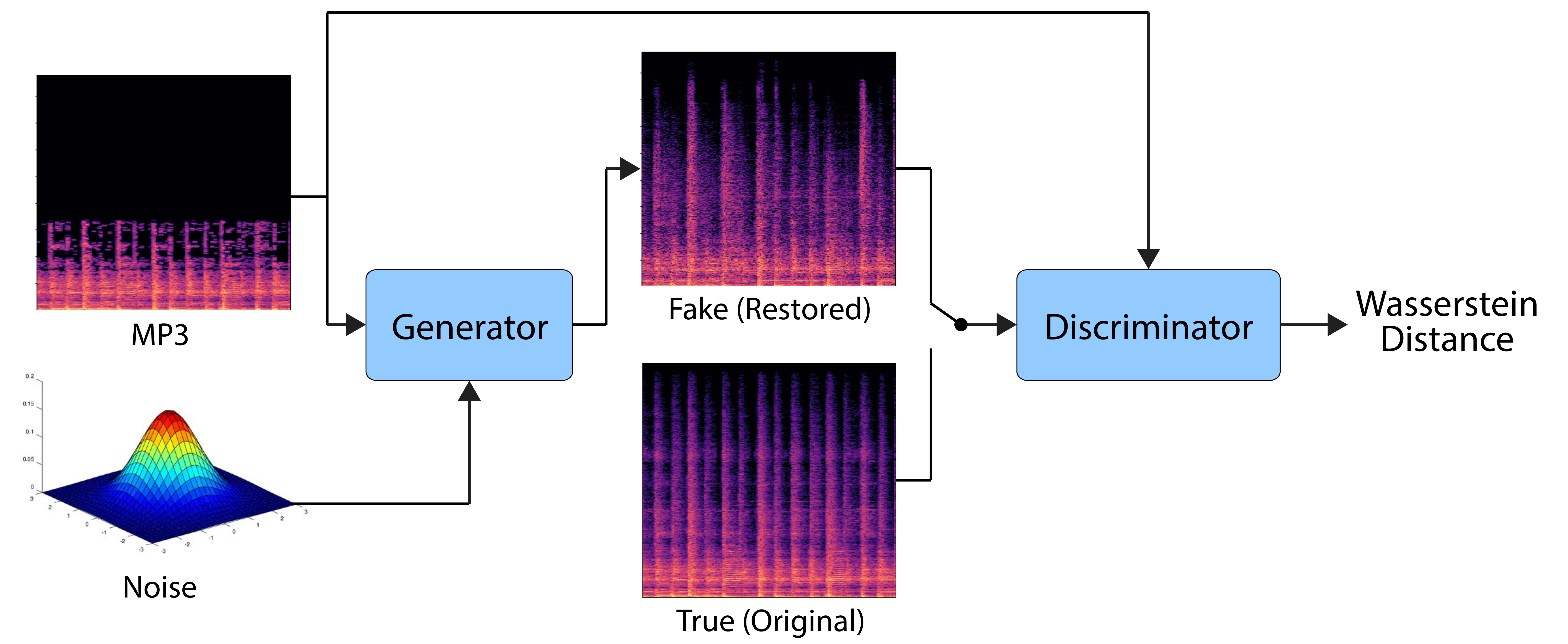}
\caption{Schematic depiction of the architecture and training procedure.}
\label{fig:training}
\end{figure}
\begin{paracol}{2}
\switchcolumn

The model employed in this work is a Generative Adversarial Network (GAN), conditioned on spectrogram representations of MP3-compressed audio files (see Figure~\ref{fig:training} for an overview on architecture and training).
As common in GANs, there are two separate models, the generator $G$ and the critic $D$.
$G$ receives as input an excerpt of an MP3-compressed musical audio signal in spectrogram representation ${y}$
(i.e., non-linearly scaled complex STFT components, see Section \ref{sec:signed_sqrt})
and learns to output a restored version ${\hat{x}}$ of that excerpt (i.e., the fake data), approximating the original, high-quality signal ${x}$.
$D$ learns to distinguish between such restorations ${\hat{x}}$ and original high-quality versions of the signal ${x}$ (i.e., the true data).
In addition to the true/fake data, $D$ also receives the MP3 versions of the respective excerpts. That way, it is ensured that the information present in the MP3 data is faithfully preserved in the output of $G$.
We test stochastic and deterministic generators in our experiments. 
For the stochastic models, we also provide some noise input ${z} \sim \mathcal{N}(0,\mathbf{I})$, resulting in different restorations for a given MP3 input.

As the training criterion, we use the GAN Wasserstein loss \cite{pmlr-v70-arjovsky17a} as 

\begin{equation}
    \label{eq:gan}
    \begin{split}
        \Gamma(D,G) = \frac{1}{N} \sum_i D({y}_i, {x}_i) - D({y}_i, G({y}_i, {z}_{i})),
    \end{split}
\end{equation}
and we are interested in $\min_{G} \max_{D} \Gamma(D,G)$, meaning the parameters of $G$ are optimized to minimize this loss, and the parameters of $D$ are optimized to maximize it. Note that the optimization of $G$ only affects the second term of Equation \ref{eq:gan}, resulting in a maximization of $D({y}_i, G({y}_i, {z}_{i}))$.

\subsubsection{Architecture Details}

\end{paracol}
\begin{table}[]
\begin{tabular}{lrrrrrrr}
\toprule
    Layer & In Maps & Out Maps & Kernel Size & Dilation & Padding & Non-linearity & Output Size \\
\midrule
 \texttt{Input} & - & - & - & - & - &  - & $2 \times 1024 \times (336) [212]$ \\
 \texttt{Conv1} & $2$ & $18$ & $3\times3$ &  $1$ & $1, 1$ &  PReLU &  $18 \times 1024 \times (336)[212]$ \\
 \texttt{Conv2} & $18$ & $38$ & $3\times3$ &  $2$ & $2, 2$ &  PReLU &  $38 \times 1024 \times (336)[212]$ \\
 \texttt{Conv3} & $38$ & $38$ & $3\times3$ &  $4$ & $4, 4$ & PReLU &  $38 \times 1024 \times (336)[212]$ \\
\midrule
\texttt{Conv4} & $38$ & $4096$ & $1024\times1$ &  $1$ & $0, 0$ & PReLU &  $4096 \times 1 \times (336)[212]$ \\
\texttt{Reshape1} && - & - &  - &  - & - & $128 \times 32 \times (336)[212]$ \\
\texttt{ReMap} & $128$ & $256$ & $1\times1$ &  $1$ & $0, 0$ & PReLU &  $256 \times 32 \times (336)[212]$ \\
\midrule
\texttt{Conv5} & $256$ & $256$ & $3\times3$ & $1$ & $1, (0)[1]$ & PReLU &  $256 \times 32 \times (334) [212]$ \\
\texttt{(Noise} &  &  &  &   &   &   & \\
\texttt{Concat)} & - & - & - &  - & - & - & $320 \times 32 \times 334$ \\
\midrule
\texttt{Conv6} & $(320)[256]$ & $256$ & $3\times3$ &  $2$ & $2, (0)[2]$ &  PReLU &  $256 \times 32 \times (330)[212]$ \\
\texttt{SelfGating} & - & - & - &  - & - & - &  $128 \times 32 \times (330)[212]$ \\
\texttt{Conv7} & $128$ & $256$ & $3\times3$ & $4$ & $4, (0)[4]$ & PReLU &  $256 \times 32 \times (322)[212]$ \\
\texttt{SelfGating} & - & - & - &  - & - & - &  $128 \times 32 \times (322)[212]$ \\
\texttt{Conv8} & $128$ & $256$ & $3\times3$ & $8$ & $8, (0)[8]$ & PReLU &  $256 \times 32 \times (306)[212]$ \\
\texttt{SelfGating} & - & - & - &  - & - & - &  $128 \times 32 \times (306)[212]$ \\
\texttt{Conv9} & $128$ & $256$ & $3\times3$ & $16$ & $16, (0)[16]$ &  PReLU &  $256 \times 32 \times (274)[212]$ \\
\texttt{SelfGating} & - & - & - &  - & - & - & $128 \times 32 \times (274)[212]$ \\
\texttt{Conv10} & $128$ & $256$ & $3\times3$ & $1$ & $1, (0)[1]$ & PReLU & $256 \times 32 \times (272)[212]$ \\
\texttt{SelfGating} & - & - & - &  - & - & - & $128 \times 32 \times (272)[212]$ \\
\texttt{Conv11} & $128$ & $256$ & $3\times3$ & $2$ & $2, (0)[2]$ & PReLU & $256 \times 32 \times (268)[212]$ \\
\texttt{SelfGating} & - & - & - &  - & - & - & $128 \times 32 \times (268)[212]$ \\
\texttt{Conv12} & $128$ & $256$ & $3\times3$ & $4$ & $4, (0)[4]$ &  PReLU &  $256 \times 32 \times (260)[212]$ \\
\texttt{SelfGating} & - & - & - &  - &  - & - & $128 \times 32 \times (260)[212]$ \\
\texttt{Conv13} & $128$ & $256$ & $3\times3$ & $8$ & $8, (0)[8]$ & PReLU & $256 \times 32 \times (244)[212]$ \\
\texttt{SelfGating} & - & - & - &  - & - & - &  $128 \times 32 \times (244)[212]$ \\
\texttt{Conv14} & $128$ & $256$ & $3\times3$ & $16$ & $16, (0)[16]$ & PReLU & $256 \times 32 \times (212)[212]$ \\
\texttt{SelfGating} & - & - & - &  - &  - & - & $128 \times 32 \times 212$ \\
\midrule
\texttt{(Reshape2)} && - & - & - & - & - &  $4096 \times 1 \times 212$ \\
\texttt{(DeConv4)} & $38$ & $4096$ & $1024\times1$ & $1$ & $0, 0$ &  PReLU &  $38 \times 1024 \times 212$ \\
\midrule
\texttt{(DeConv3)} & $38$ & $38$ & $3\times3$ & $4$ & $4, 4$ & PReLU & $38 \times 1024 \times 212$ \\
\texttt{(DeConv2)} & $18$ & $38$ & $3\times3$ & $2$ & $2, 2$ & PReLU & $18 \times 1024 \times 212$ \\
\texttt{(DeConv1)} & $2$ & $18$ & $3\times3$ & $1$ & $1, 1$ & PReLU &  $2 \times 1024 \times 212$ \\
\texttt{(Output)} && - & - & - &  - & - & $2 \times 1024 \times 212$ \\
\midrule
\texttt{[Conv15]} & $128$ & $256$ & $3\times3$ & $1$ & $1, 1$ & PReLU & $256 \times 32 \times 212$ \\
\texttt{[Conv16]} & $256$ & $1$ & $32\times1$ &  $1$ & $0, 0$ & - & $1 \times 1 \times 212$ \\

\bottomrule
\end{tabular}
\hspace{-5cm}
\caption{Architecture details of generator $G$ and critic $D$ for $4$-second-long excerpts (i.e., 336 spectrogram frames), where $(\cdot)$-brackets mark information applying only to $G$, and information in $[\cdot]$-brackets applies only to $D$. During training, no padding is used in the time dimension for $G$ resulting in a shrinking of its output to $212$ time steps.}
\label{tab:architecture}
\end{table}
\begin{paracol}{2}
\switchcolumn

For details on the implemented architecture, please refer to Table \ref{tab:architecture}.
We implement both the generator $G$ and the critic $D$ convolutional in time.
This allows us to use less overlap (i.e., $50\%$) when chopping up the training data, as the convolutional architectures obtain differently shifted versions of the input by design.
At test time, $G$ is applied to variable-length and potentially relatively long input sequences (e.g., full songs).
In such a setting, $G$ does not perform very well if trained on short excerpts with zero-padding in the time dimension.
Therefore, we do not use zero-padding for $G$ during training.

The critic $D$ is convolutional in time, too, resulting in as many loss outputs as there are spectrogram frames in the input (the individual costs are simply averaged for computing the final Wasserstein loss).
We are using two convolutional groups throughout the critic stack, which amounts to two independent critics.
Only in the output layer, those two groups are joined again.
This is to provide $D$ with two different views on the input, (1) the signed square-root of the complex STFT components (see Section \ref{sec:signed_sqrt}) and (2) the \emph{square-root} of the magnitude spectrum of the generator output (we found empirically that this resulted in more stable training than when using the log-magnitude spectrogram).

Many convolutional architectures with the same output and input size used for data restoration and in-painting employ the symmetrical U-Net paradigm (first introduced in \cite{DBLP:conf/cvpr/LongSD15}) with bottleneck layers and skip connections.
In contrast, the architecture proposed in this work is non-symmetrical, mainly facilitating dilated convolutions for increasing receptive fields, and the main part of the architectures of $G$ and $D$ are identical (see Table \ref{tab:architecture}).
Only the top parts of the stacks differ, wherein $G$ the aggregated information is fed into deconvolution layers, while in $D$ the information is used to compute the Wasserstein distance.

We use Parametric Rectified Linear Units (PReLUs) \cite{DBLP:conf/iccv/HeZRS15} for all layers, and skip connections for $D$ in the convolutional layers \texttt{Conv6} - \texttt{Conv14} in Table \ref{tab:architecture}. The noise input $z$ (to the generator $G$) is simply repeated in the two convolutional dimensions and concatenated to layer \texttt{Conv5}.

\subsubsection{Gated Convolutions}\label{sec:gating}
In order to increase the architecture size under limited resources, a handy modification to common convolutions are self-gating convolutional layers.
This idea was also proposed in \cite{DBLP:conf/icml/DauphinFAG17}, but we are using PReLU activations instead of linear units for the gated output units (linear units resulted in unstable training).
The characteristic of (self-)gating convolutions is that half the output maps of each convolutional layer are used to element-wise gate the other half of the output maps (where we use sigmoid non-linearities on the gating units and PReLUs on the gated units). We found that with self-gating layers, the network's performance does not degrade, even though the operation effectively halves the number of output maps for each layer. The advantage of self-gating is a considerable reduction of memory, as the successive layer receives only half the input maps compared to non-self-gating layers. In practice, we use only one layer with twice as many output maps as input maps, which are then used for self-gating. Formally, we describe the operation with two different weight matrices as

\begin{equation}
    y_l = \text{PReLU}_l(x_{l} * W_l + a_l) \cdot \sigma(x_{l} * V_l + b_l),
\end{equation}
where $x_l \in \mathbb{R}^{R \times P \times T}$ is the input to layer $l$ with $R$ convolutional input maps, $y_l \in \mathbb{R}^{S \times P \times T}$ is the resulting self-gated output of layer $l$ with $S$ convolutional output maps (where $S = R$ in our architecture), $W_l, V_l \in \mathbb{R}^{R \times S \times K \times K}$ are two weight matrices with quadratic kernels of size $K \times K$, and $a_l, b_l \in \mathbb{R}^{S}$ are bias vectors. $\text{PReLU}$ are parametric ReLU activations \cite{DBLP:conf/iccv/HeZRS15}, $\sigma$ is the sigmoid non-linearity, $*$ is the convolution operator and $\cdot$ is the Hadamard product.
This operation is applied to all convolutional layers \texttt{Conv6} - \texttt{Conv14} in Table \ref{tab:architecture}.

\subsubsection{Frequency Aggregation Filters}\label{sec:freq_aggregation}

\end{paracol}
\begin{figure}
\widefigure
\includegraphics[width=.8\textwidth]{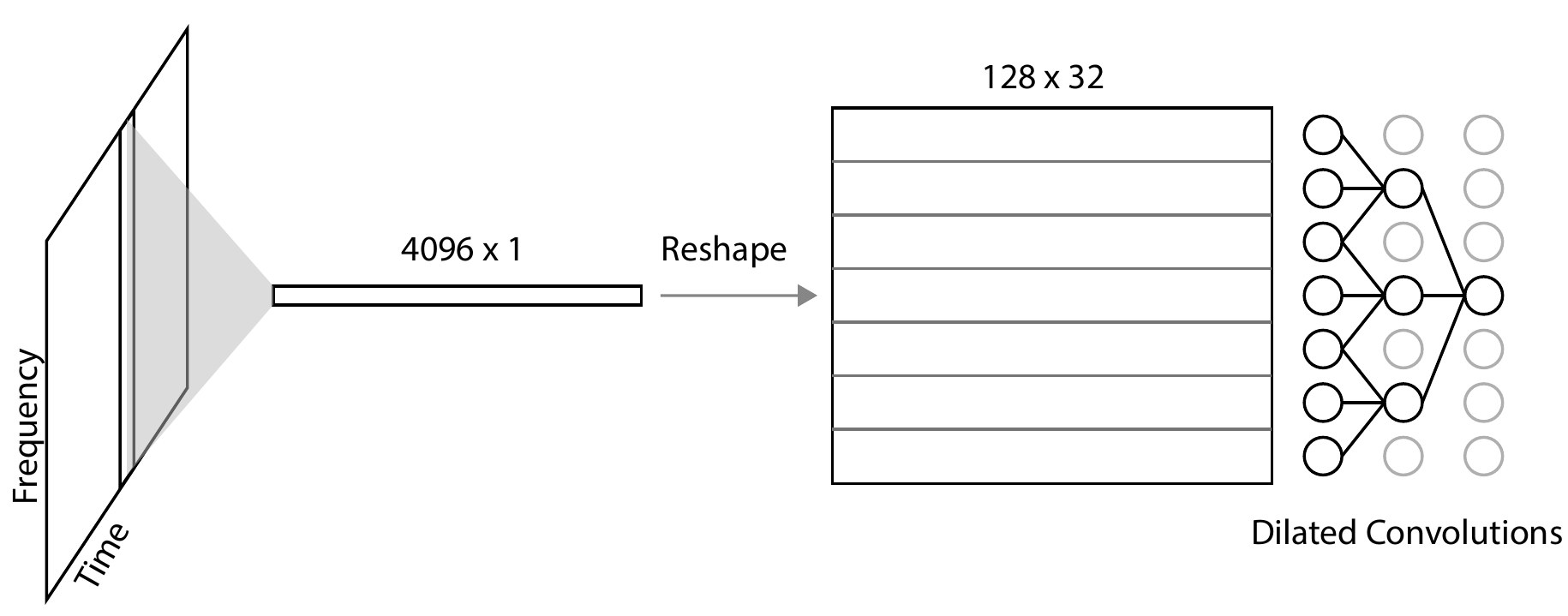}
\caption{Schematic depiction of the Frequency Aggregation module for \emph{one time frame}. The frame is filtered with $4096$ kernels resulting in filter responses of size $4096 \times 1$ (\texttt{Conv4} in Table \ref{tab:architecture}). By reshaping, the responses are then separated into 32 groups (of size 128 each) and re-combined again through a stack of dilated convolutions (\texttt{Conv5} - \texttt{Conv14} in Table \ref{tab:architecture}).}
\label{fig:freq_aggregation}
\end{figure}
\begin{paracol}{2}
\switchcolumn

The neural network architectures used in audio processing are often derived from the visual domain.
Convolutional neural networks are particularly well-suited for image processing because, in natural images, close pixels are usually higher correlated than pixels further apart.
Using stacks of convolutional layers, the filter kernels in the lower layers can learn highly correlated information, and filters in the higher layers can learn more complex combinations of filter responses of the lower layers.
That way, as a rule of thumb, the higher up in the convolutional hierarchy, the less correlated information is represented, which results in a hierarchical aggregation of pixel information that is well-suited for natural images.

Such a correlation assumption may also hold for the time dimension when working with musical audio data in spectrogram representation.
However, it does not hold in the frequency dimension, where highly correlated spectral energy is potentially spread over the whole frequency range.
In order to comply with this characteristic, it is common to employ non-rectangular filters in the input layer of a convolutional network stack, for example, kernels of shape [1, 32] in \cite{pons2019deep}.
Still, when considering harmonics of tonal instruments or percussive sounds, correlated information may be so distant in the frequency axis that also with vertical filter kernels, a complete acoustic source may only be fully represented in the highest layer of the hierarchy.
This fact contradicts the useful characteristic of efficient aggregation of information in convolutional network stacks to represent the least correlated information (i.e., the most complex patterns) in the highest layers of the hierarchy.

In order to tackle the problem of highly correlated frequency bins very distant in the frequency dimension, we take a novel approach.
As it is not obvious before training which frequency bins are most correlated in the training data, and it is therefore not clear how to best design the architecture, we allow the network to \emph{learn} a useful hierarchy of frequency aggregation during training (see Figure \ref{fig:freq_aggregation}).
To that end, in Layer \texttt{Conv4} (see Table \ref{tab:architecture}), we use $4096$ filter kernels that span the whole frequency dimension and only convolve in time (i.e., no padding in the frequency dimension).
Then, we reshape the output maps (see \texttt{Reshape1} in Table \ref{tab:architecture}) so that we again obtain a 2D convolutional architecture and let the network learn which filter kernels are most correlated, i.e., in what layer of the hierarchy which filter responses need to be brought together (throughout layers \texttt{Conv5} - \texttt{Conv14}).
In the generator $G$, \texttt{Reshape2} reshapes back to $4096$ feature maps and \texttt{DeConv4} inverts the frequency aggregation.

\subsection{Training Procedure}\label{sec:training_procedure}
Each model is trained for $40$k iterations and a batch size of $12$, which takes about 2 days on two NVIDIA Titan RTX with 24GB memory each. We use the ADAM optimizer \cite{DBLP:journals/corr/KingmaB14} with a learning rate of $1$e-$3$ and gradient penalty loss, to restrict the gradients of $D$ to $1$ Lipschitz \cite{DBLP:conf/nips/GulrajaniAADC17}. We also use a loss term that penalizes the magnitudes of the output of $D$ for \emph{real} input data, preventing the loss from drifting. Furthermore, He's initialization is performed for all the layers in the architecture \cite{DBLP:conf/iccv/HeZRS15}.

\subsubsection{Preventing Mode Collapse}\label{sec:mode_collapse}
We found that the Generator $G$ tends to ignore the input noise ${z}$.
This may be because $G$ is densely conditioned on ${x}$, and the output variability in the data is limited given an input with a specific characteristic.
In order to prevent such a mode collapse, for updating $G$ during training, we define an additional cost term which is maximal when the noise input to $G$ does not influence the output of $G$.
To that end, for a fixed ${y}_i$, we compute the ratio between the Euclidean distance of two arbitrary input noise vectors $\{{z}_i, {z}_j\}$, and the distances between the corresponding frequency profiles (summing over the time axis) and the rhythm profiles (summing over the frequency axis) of the output magnitude spectrogram of G, resulting in loss $\mathcal{L}_{\scriptstyle\text{freq}}$ and $\mathcal{L}_{\scriptstyle\text{rhyt}}$, respectively:


\begin{equation}
    \mathcal{L}_{\scriptstyle\text{profile}} = \frac{\vartheta \, {\lVert {z}_i - {z}_j \rVert}}
    {{\left\lVert d^{{}^\intercal}{\hat{P}_{z_i}^{\circ \frac{1}{2}}} - {d^{{}^\intercal}{\hat{P}_{z_j}^{\circ \frac{1}{2}}}} \right\rVert_{\scriptstyle p}^{\scriptstyle p}}},
\end{equation}
where $\hat{P}_{z_i}^{\circ \frac{1}{2}}$ is the Hadamard root of the power spectrum of the output of $G$ for input noise vector ${z_i}$, $d$ is a column vector of 1s for the frequency profiles (resulting in the loss $\mathcal{L}_{\scriptstyle\text{freq}}$) and a row vector of 1s for the rhythm profiles (resulting in $\mathcal{L}_{\scriptstyle\text{rhyt}}$). The scalar $\vartheta$ controls the strength of the regularization, $p=1.3$ for frequency profiles and $p=1.6$ for rhythm profiles in our experiments. 

In practice, for each conditional input ${y}_i$ to $G$ (i.e., each instance with index $i$ in a batch), we compute two outputs $G({x}_k, \{{z}_i, {z}_j\}_k)$ using randomly sampled $\{{z}_i, {z}_j\}_k \sim \mathcal{N}(0,{I})$, and use those outputs to compute $\mathcal{L}_{\text{profile}}$, as well as the common gradient update of $G$.
Note that in order to minimize $\mathcal{L}_{\text{profile}}$, $G$ could simply learn to introduce huge changes in its output when the input noise ${z}$ changes. However, in practice, this is prevented by the Wasserstein loss, which introduces a strong bias towards plausible (i.e., obeying the data distribution)  outputs of $G$. Therefore, $\mathcal{L}_{\text{profile}}$ is effective in pushing the generations of $G$ away from deterministic outputs while the overall training process remains stable.

\subsection{Data}\label{sec:data}
The model is trained on pairs of audio data, where one part is the MP3 version, the other part is a high-quality (44.1 kHz) version of the signal. We use a dataset of approximately 64 hours of Nr 1 songs of the US charts between 1950 and 2020. The high-quality data is then compressed to $16$kbit/s, $32$kbit/s and $64$kbit/s mono MP3 using the LAME MP3 codec, version 3.100.\footnote{\url{https://lame.sourceforge.io/} (accessed on 31 May 2021)}
The total number of songs is first divided into train, eval, and test sub-sets with a ratio of 80\%, 10\%, 10\%, respectively. We then split each of the songs into $4$-second-long segments with $50\%$ overlap for training and validation. For the subjective evaluation (see Section \ref{sec:mos}), we split the songs into segments of $8$ seconds.

\subsubsection{Data Representation}\label{sec:signed_sqrt}
The main representation used in the proposed method are the complex STFT components of the audio data $h_{j,k} \in \mathbb{C}^{JK}$, as it has been shown that this representation works well for audio generation with GANs in \cite{DBLP:conf/eusipco/NistalLR20}. The STFT is computed with window size 2048, and a hop size of 512. In addition, we perform non-linear scaling to all complex components, in order to obtain a scaling which is closer to human perception than when using the STFT components directly. This is, we transform each complex STFT coefficient $h_{j,k} = a_{j,k} + i\:b_{j,k}$ by taking the signed square-root of each of its components $h^{\sigma}_{j,k} = \sigma(a_{j,k}) + i\:\sigma(b_{j,k})$, where the signed square-root is defined as

\begin{equation}
    \sigma(r) = \sign(r) \sqrt{|r|}.
\end{equation}

\subsection{Evaluation}\label{sec:evaluation}

We perform objective and subjective evaluations for the proposed method.
The main goal of the evaluation is to assess the similarity between the reference signals (i.e., the high-quality signals) and the signal approximations (i.e., MP3 versions of the audio excerpts or outputs of the proposed model).
The objective metrics used include Log-Spectral Distance (LSD), Mean Squared Error (MSE), Signal-to-Noise Ratio (SNR), Objective Difference Grade (ODG), and Distortion Index (DI).
We also perform a subjective evaluation in the form of the Mean Opinion Score (MOS), which is described in Section \ref{sec:mos}.

\subsubsection{Objective Difference Grade and Distortion Index}
The Objective Difference Grade (ODG) is a computational approximation to subjective evaluations (i.e., the \emph{subjective difference grade}) of users when comparing two signals.
It ranges from $0$ to $-4$, where lower values denote worse similarities between the signals.
The Distortion Index (DI) is a metric that is differently scaled but correlated to the ODG and can be seen as the amount of distortion between two signals.
Both the ODG and DI are based on a highly non-linear psychoacoustic model, including filtering and masking to approximate the human auditory perception.
They are part of the Perceptual Evaluation of Audio Quality (PEAQ) ITU-R recommendation (BS.1387-1, last updated 2001) \cite{thiede2000peaq}.
We use an openly available implementation of the basic version (as defined in the ITU recommendation) of PEAQ\footnote{\url{https://github.com/akinori-ito/peaqb-fast} (accessed on 31 May 2021)}, including ODG and Distortion Index (DI).
Even though PEAQ was initially designed for evaluating audio codecs with \emph{minimal} coding artifacts, we found that the results correlate well with our perception.

\subsubsection{Log-Spectral Distance}\label{sec:lsd}
The log-spectral distance (LSD) is the Euclidean distance between the log-spectra of two signals and is invariant to phase information.
Here, we calculate the LSD between the spectrogram of the reference signal and that of the signal approximation.
This results in the equation
\begin{equation}
    LSD=\frac{1}{L} \sum_{l=0}^{L-1}\sqrt{\frac{1}{W} \sum_{f=0}^{W-1} \left[ 10\log_{10} \frac{P(l,f)}{\hat{P}(l,f)} \right]^2},
\end{equation}
where $P$ and $\hat{P}$ are the power spectra of $x$ and $\hat{x}$, respectively, $L$ is the total number of frames, and $W$ is the total number of frequency bins.

\subsubsection{Mean Squared Error}
The LSD described above (see Section \ref{sec:lsd}) is particularly high when comparing MP3 data with high-quality audio data.
This is because it is standard practice found in many MP3 encoders (including the one we use) to perform a high-cut, removing most frequencies above a specific cut-off frequency.
For values close to zero, a log-scaling introduces negative numbers with very high magnitudes.
Therefore, when comparing log-scaled power spectra of MP3 and PCM, we obtain particularly high distances.
This generally favors algorithms that add frequencies in the upper range (like the proposed method).
In this regard, a fairer comparison is the Mean Squared Error (MSE) between the \emph{square-root} of the power spectra $P$ of the two signals:
\begin{equation}
    MSE=\frac{1}{L} \sum_{l=0}^{L-1} \frac{1}{W} \sum_{f=0}^{W-1} \left[ \sqrt{P(l,f)}-\sqrt{\hat{P}(l,f)} \right]^2.
\end{equation}

\subsubsection{Signal-to-Noise Ratio}
The signal-to-noise ratio (SNR) measures the ratio between a reference signal and the approximation residuals.
As it is computed in the \emph{time domain}, it is highly sensitive to phase information.
The SNR is calculated as
\begin{equation}
    \text{SNR}=10\log_{10} \frac{{{}\lVert s \rVert}_{\scriptscriptstyle 2}^{\scriptscriptstyle 2}}{{{}\lVert s - \hat{s} \rVert}_{\scriptscriptstyle 2}^{\scriptscriptstyle 2}},
\end{equation}
where $s$ is the reference signal, and $\hat{s}$ is the signal approximation.

\subsubsection{Mean Opinion Score}\label{sec:mos}
We ask 15 participants (mostly expert listeners) to provide absolute ratings (i.e., no reference audio excerpts) of the perceptual quality of isolated musical excerpts.
The listening test is performed with random, $8$ second-long audio excerpts of the test set. We present to the listeners $5$ high-quality audio excerpts, $15$ MP3s ($5 \times 16$kbit/s, $5 \times 32$kbit/s and $5 \times 64$kbit/s) and $50$ restored versions (using $25$ stochastic restorations with random noise $z$ and $25$ deterministic restorations). Among these $25$ restorations per model we restored $10 \times 16$kbit/s, $10 \times 32$kbit/s and $5 \times 64$kbit/s MP3s. All together this results in 70 ratings per user.

The participants were asked to give an overall quality score and instructed to consider both the extent of the audible frequency range and noticeable, annoying artifacts. They provided their rating using a Likert-scale slider with 5 quality levels (1) \emph{very bad}, 2) \emph{poor}, 3) \emph{fair}, 4) \emph{good} and 5) \emph{excellent}). From these results, we compute the Mean Opinion Score (MOS) \cite{mos}.

\section{Results and Discussion}\label{sec:results}
\end{paracol}
\begin{figure}
\hspace{-8pt}
\begin{tabular}{ccccc}
\includegraphics[trim=0 20 45 0,clip, width=0.18\textwidth]{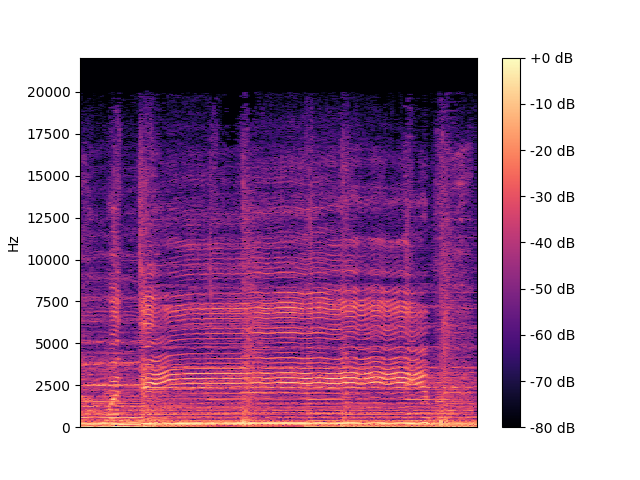} & 
\includegraphics[trim=0 20 45 0,clip, width=0.18\textwidth]{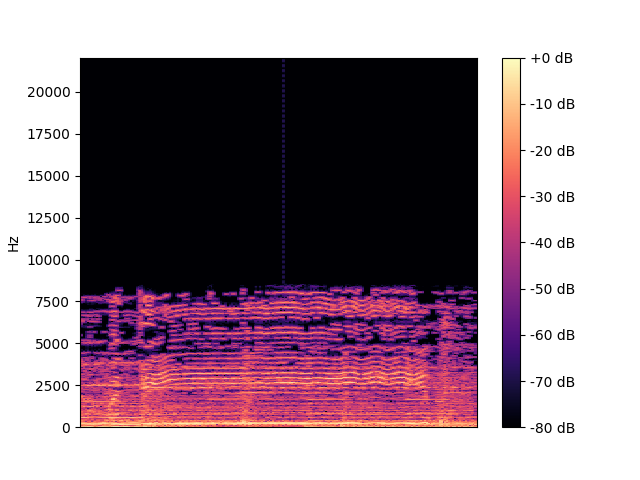} & 
\includegraphics[trim=0 20 45 0,clip, width=0.18\textwidth]{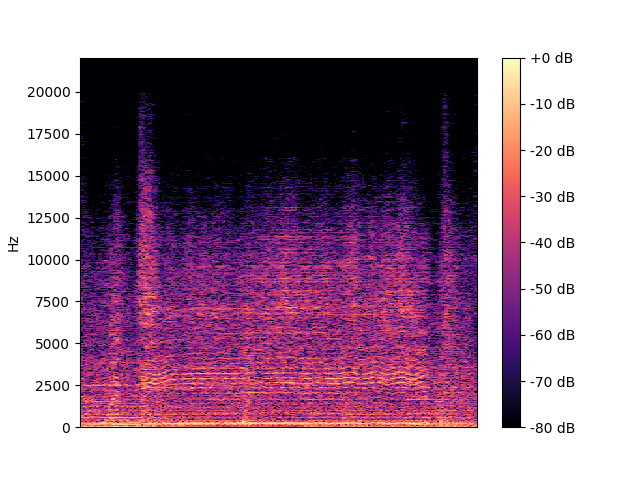} & 
\includegraphics[trim=0 20 45 0,clip, width=0.18\textwidth]{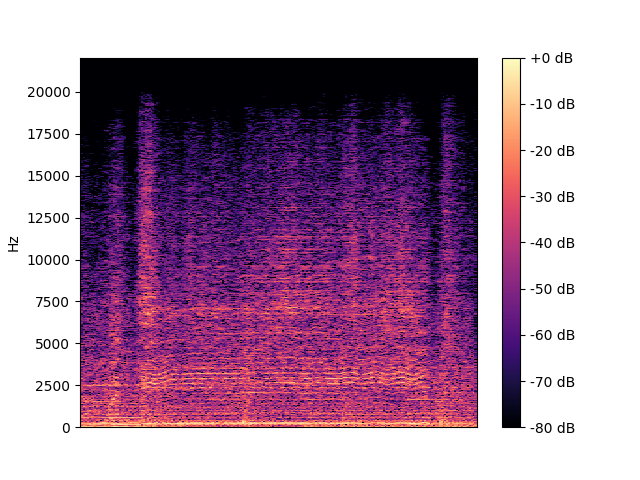} & 
\includegraphics[trim=0 20 45 0,clip, width=0.18\textwidth]{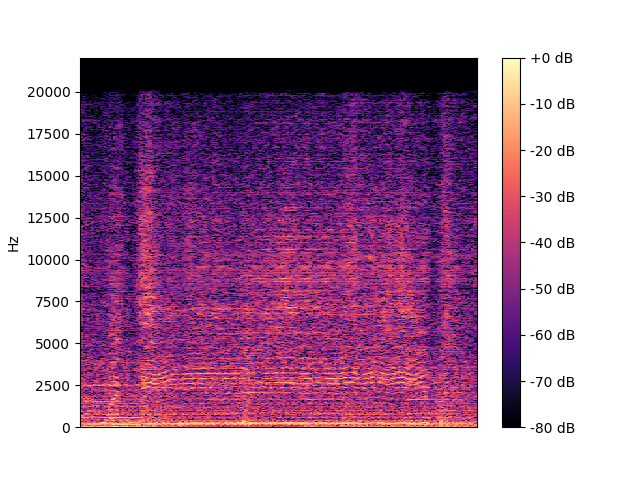} \\
\includegraphics[trim=0 20 45 0,clip, width=0.18\textwidth]{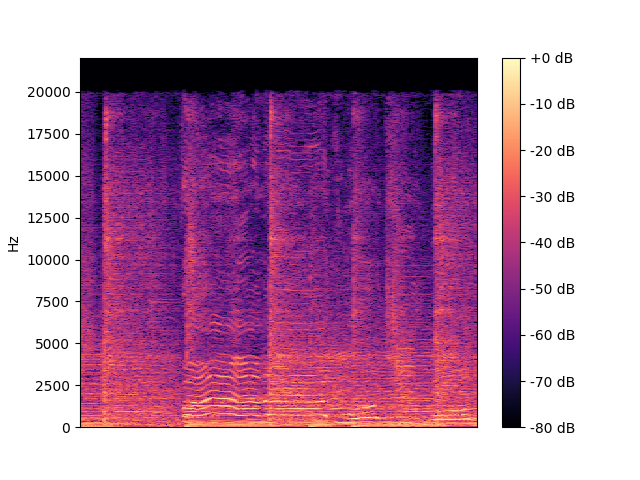} & 
\includegraphics[trim=0 20 45 0,clip, width=0.18\textwidth]{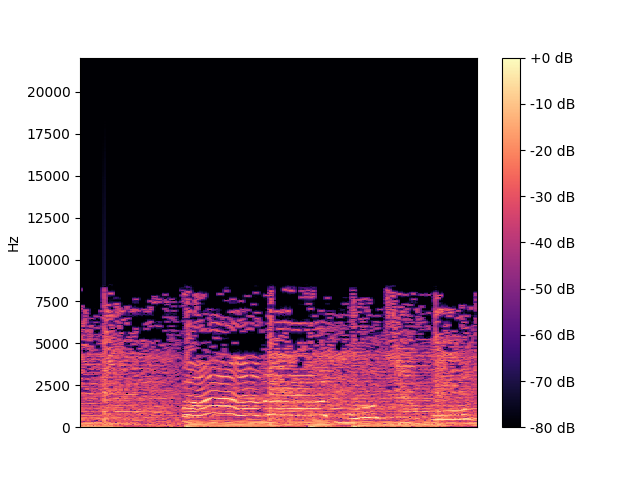} & 
\includegraphics[trim=0 20 45 0,clip, width=0.18\textwidth]{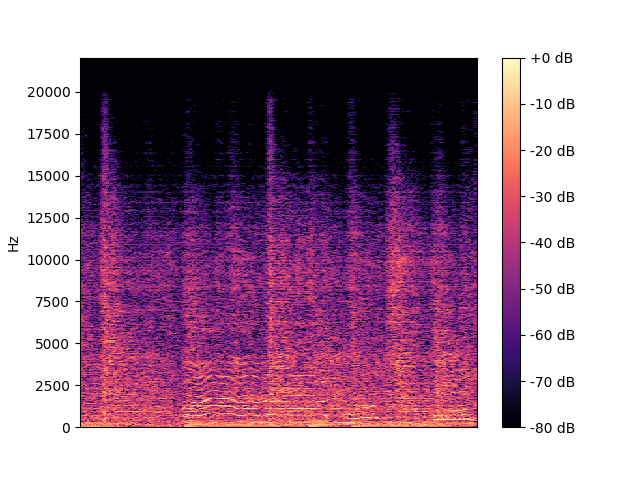} & 
\includegraphics[trim=0 20 45 0,clip, width=0.18\textwidth]{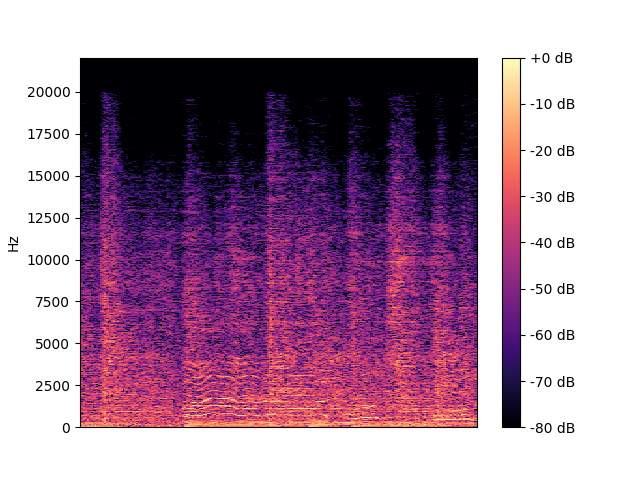} & 
\includegraphics[trim=0 20 45 0,clip, width=0.18\textwidth]{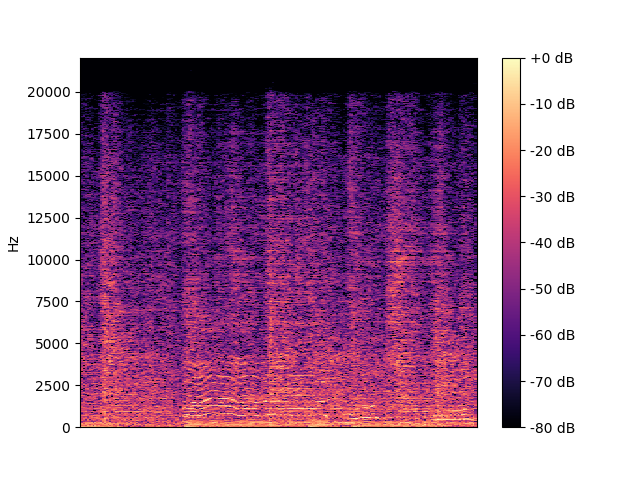} \\
\includegraphics[trim=0 20 45 0,clip, width=0.18\textwidth]{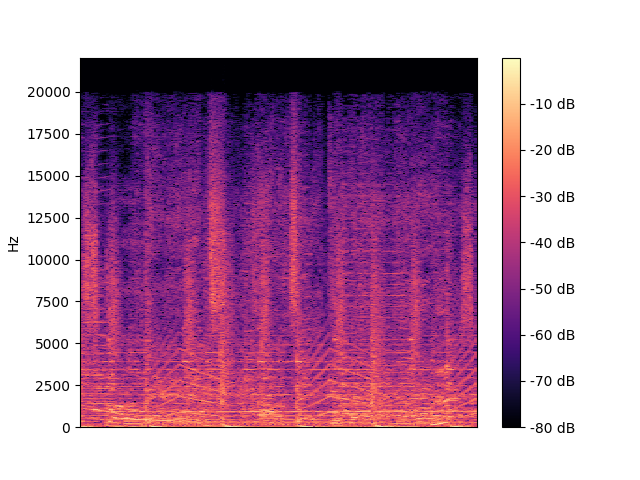} & 
\includegraphics[trim=0 20 45 0,clip, width=0.18\textwidth]{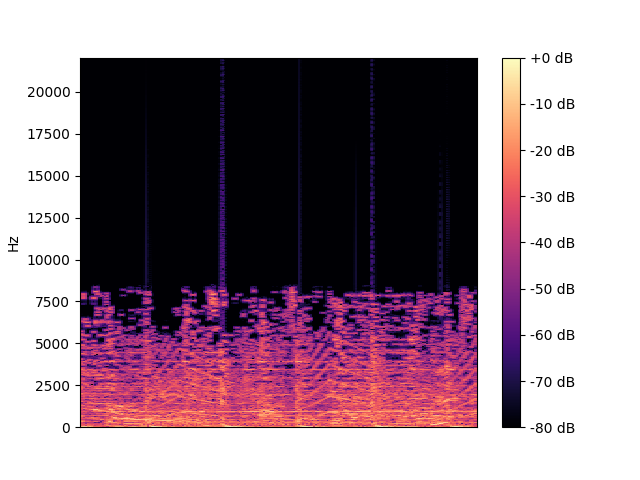} & 
\includegraphics[trim=0 20 45 0,clip, width=0.18\textwidth]{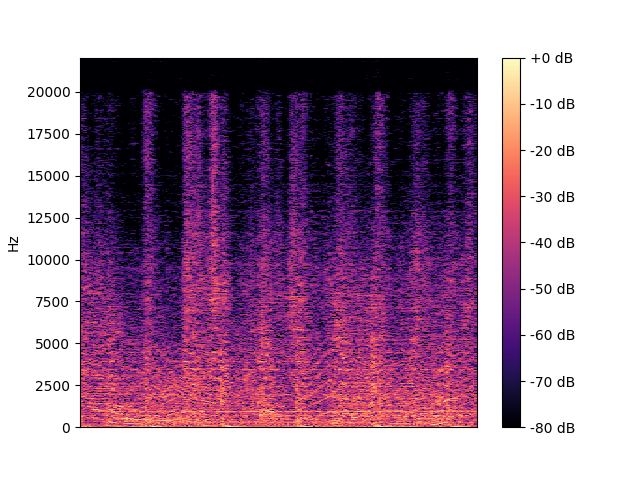} & 
\includegraphics[trim=0 20 45 0,clip, width=0.18\textwidth]{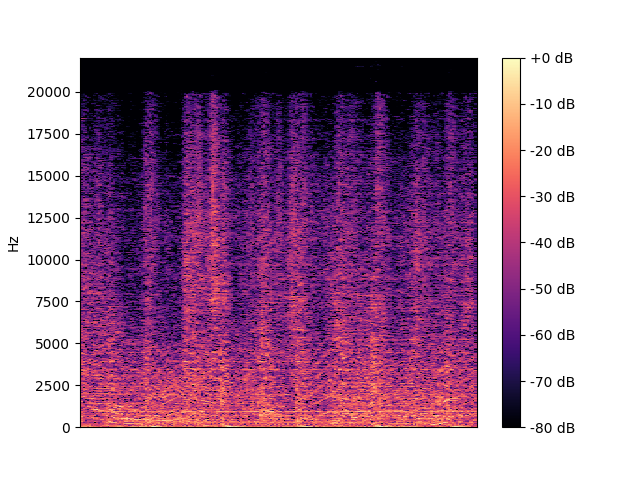} & 
\includegraphics[trim=0 20 45 0,clip, width=0.18\textwidth]{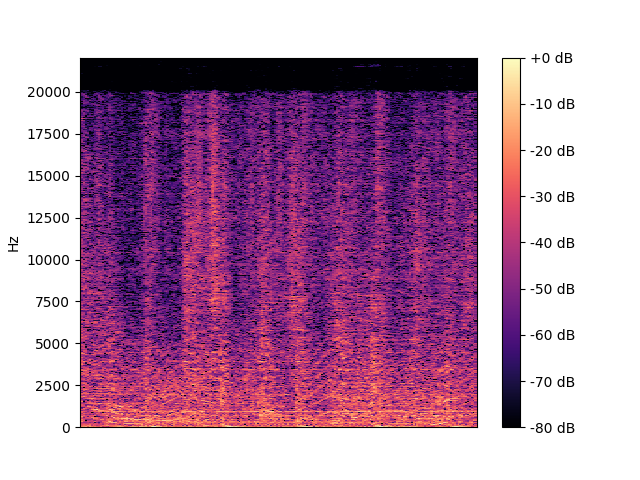} \\
(a) & (b) & (c) & (d) & (e) \\
\end{tabular}
\caption{Spectrograms of (a) original audio excerpts, (b) corresponding $32$kbit/s MP3 versions, and (c), (d), (e) restorations with different noise $z$ randomly sampled from $\mathcal{N}(0,\mathbf{I})$.}
\label{fig:spectrograms}
\end{figure}
\begin{paracol}{2}
\switchcolumn

\end{paracol}
\begin{figure}
\begin{tabular}{cc}
\includegraphics[trim=5 2 40 30,clip, width=0.42\textwidth]{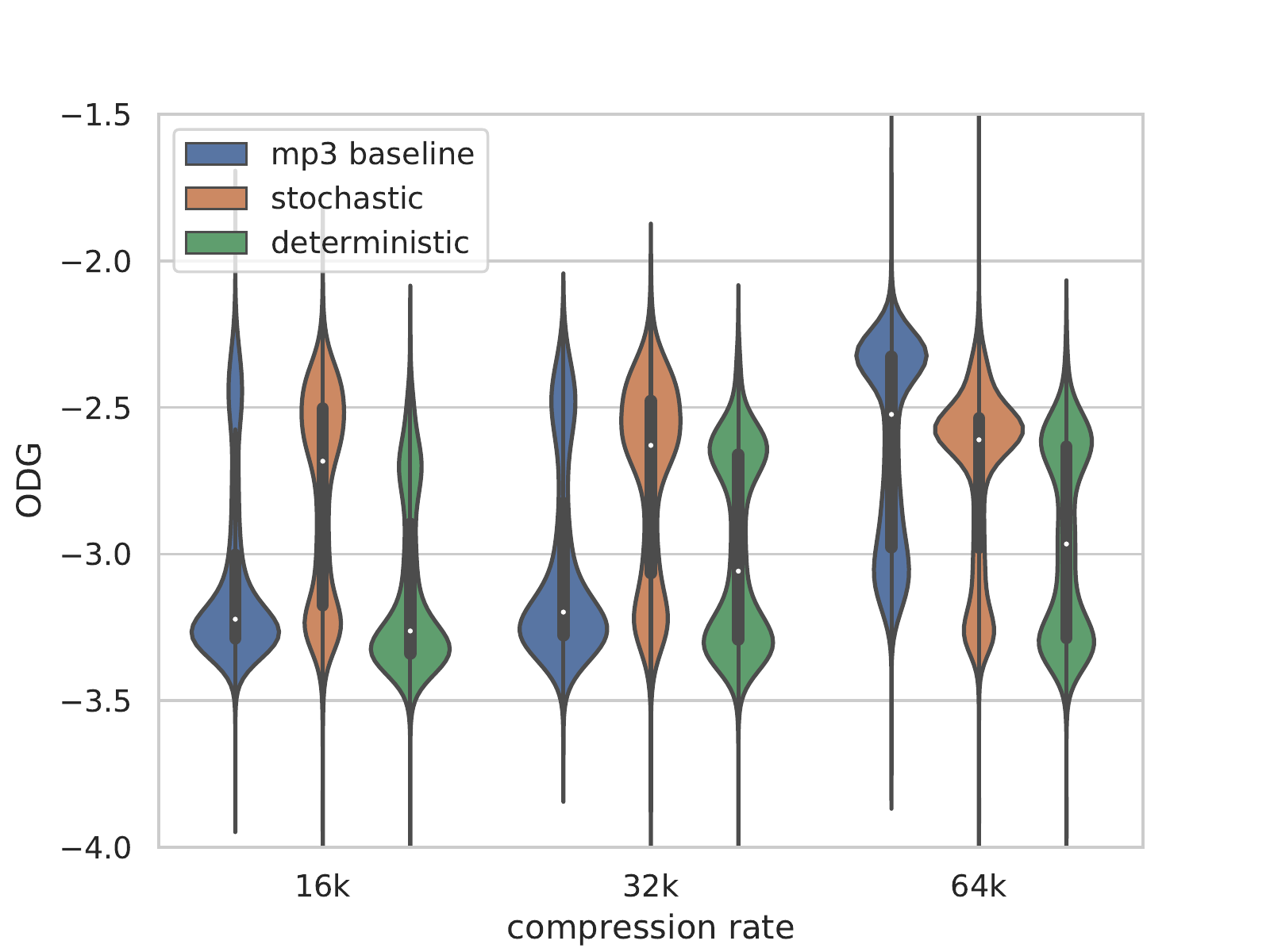} & 
\includegraphics[trim=5 2 40 30,clip, width=0.42\textwidth]{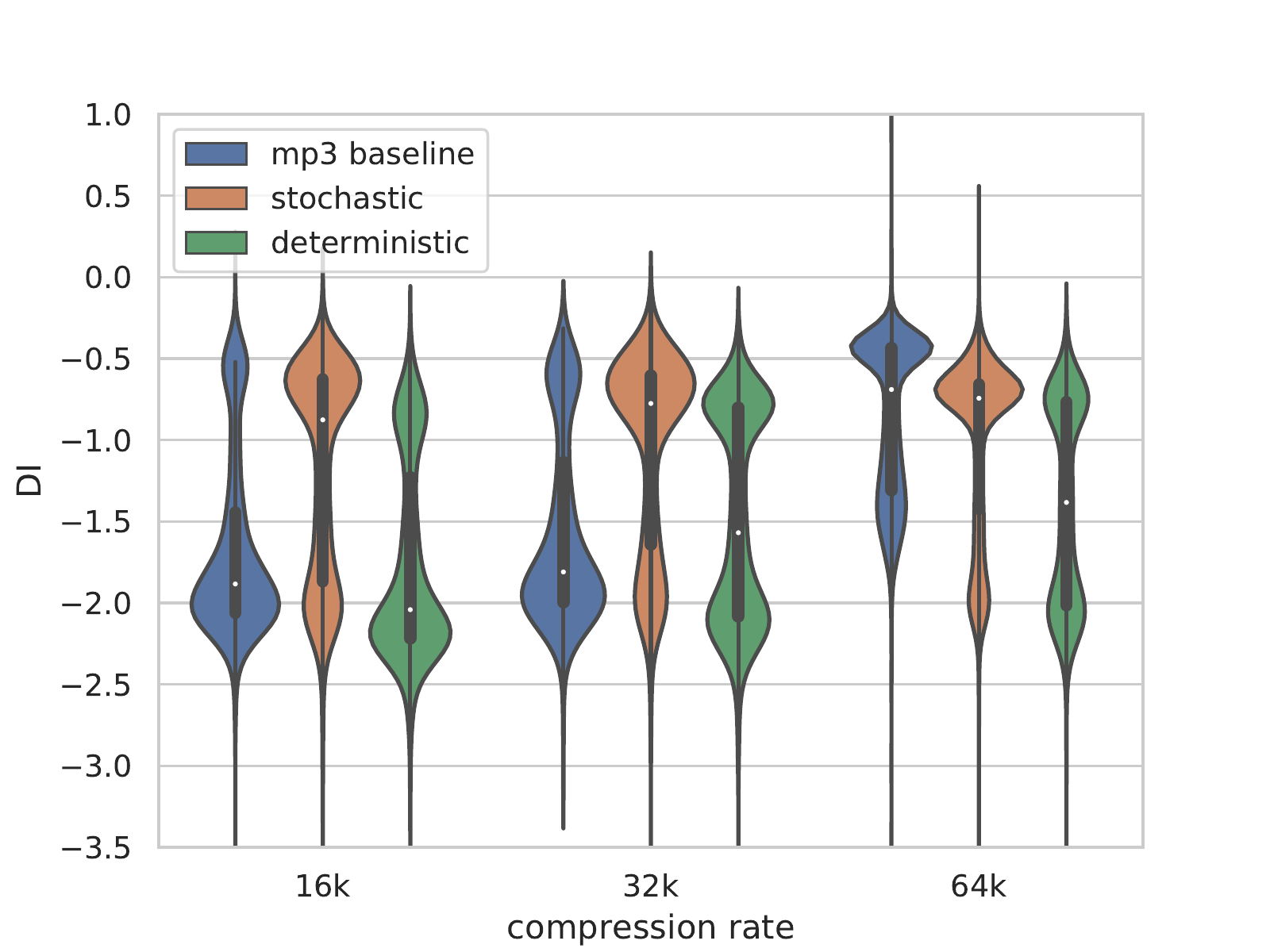} \\
\includegraphics[trim=5 2 40 30,clip, width=0.42\textwidth]{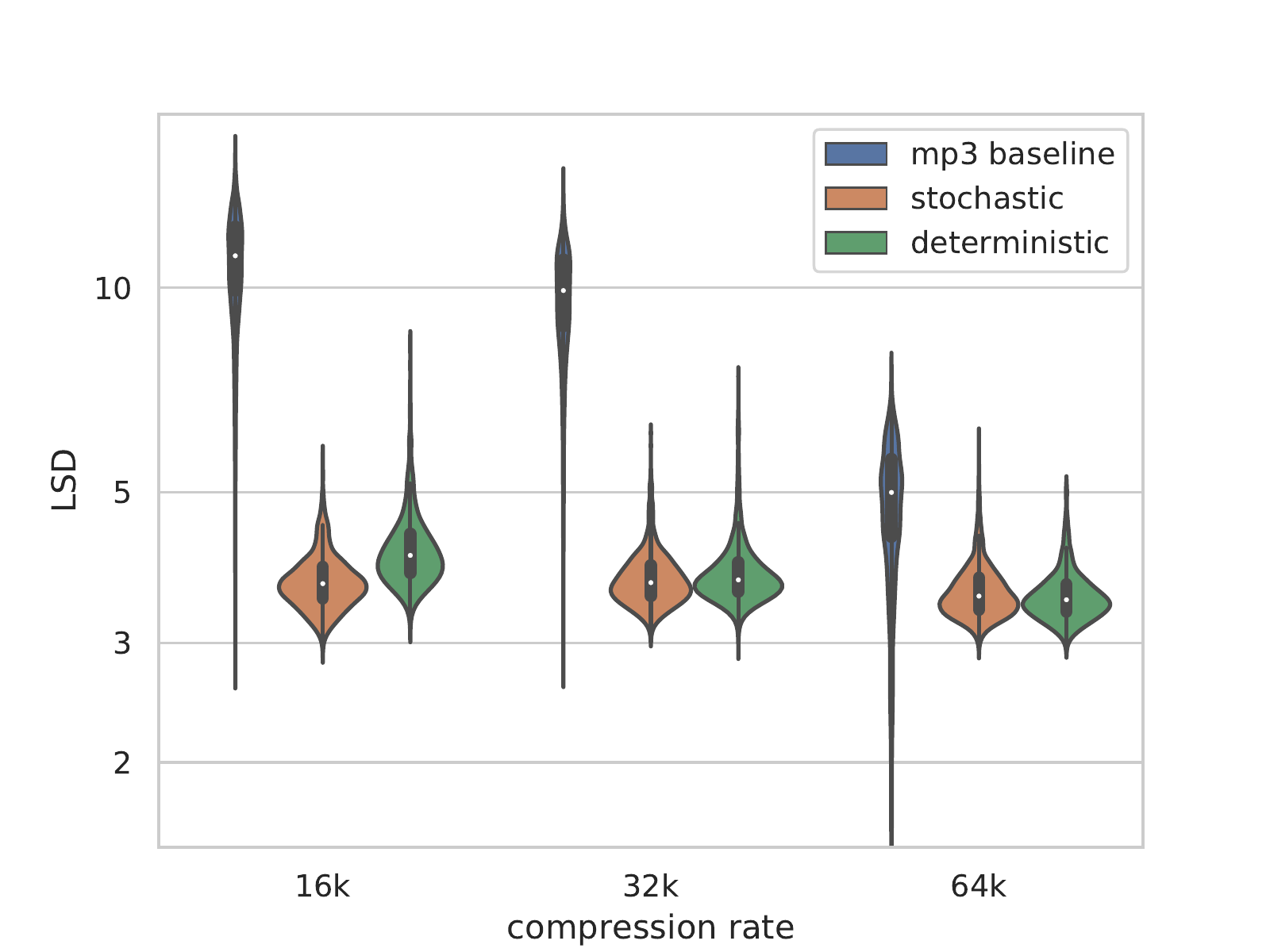} & 
\includegraphics[trim=5 2 40 30,clip, width=0.42\textwidth]{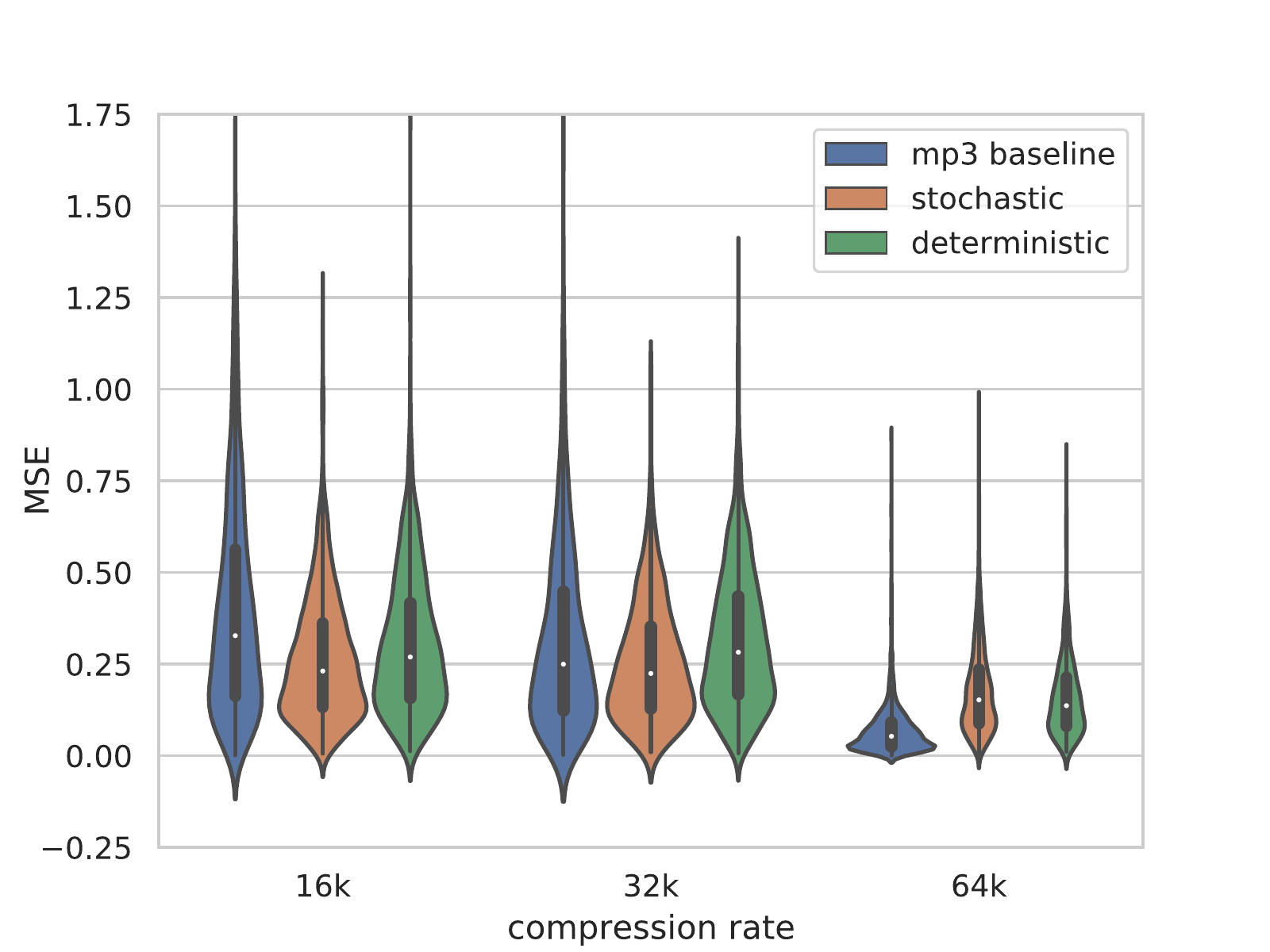} \\
\includegraphics[trim=5 2 40 30,clip, width=0.42\textwidth]{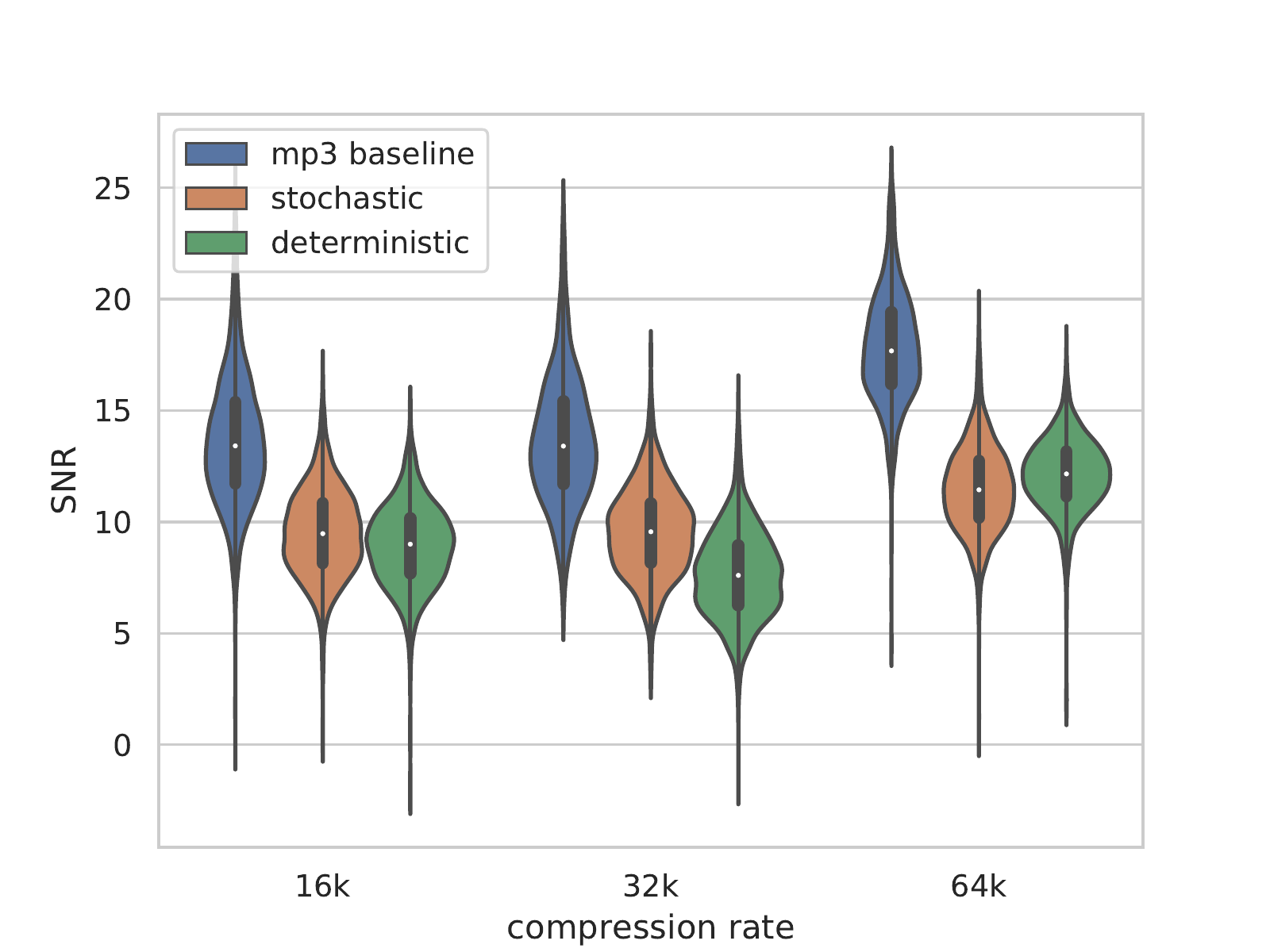} & \\
\end{tabular}
\caption{Violin plots of objective metrics for stochastic (\texttt{sto}), deterministic (\texttt{det}) models and MP3 baselines (\texttt{mp3}), for different compression rates ($16$ kbit/s, $32$kbit/s, $64$kbit/s). Higher values  are better for ODG, DI and SNR; lower values are better for LSD and MSE.}
\label{fig:box_plots}
\end{figure}
\begin{paracol}{2}
\switchcolumn

In the following, we present the results of the performed evaluations.
In Section \ref{sec:eval_obj} we discuss the results of the objective metrics and in Section \ref{sec:eval_subj} we discuss the subjective evaluation (i.e., the MUSHRA test and the MOS). Figure \ref{fig:spectrograms} provides a visual impression of the model output by comparing the spectrograms of some high-quality audio segments, the corresponding MP3 versions, and some restorations.

\subsection{Objective Evaluation}\label{sec:eval_obj}

\begin{table}[t]
\begin{center}
\begin{tabular}{lrrrrr}
\toprule
         &   ODG &    DI &   LSD &   MSE &   SNR \\
\midrule
 \texttt{mp3\_16k} & -3.08 & -1.67 & 10.98 &  0.40 & \textbf{13.69} \\
 \texttt{det\_16k} & -3.12 & -1.77 &  4.15 &  0.30 &  8.95 \\
 \texttt{sto\_16k} & \textbf{-2.80} & \textbf{-1.19} &  \textbf{3.72} &  \textbf{0.26} &  9.51 \\
\midrule
 \texttt{mp3\_32k} & -3.04 & -1.56 &  9.75 &  0.31 & \textbf{13.67} \\
 \texttt{det\_32k} & -2.99 & -1.48 &  3.83 &  0.32 &  7.66 \\
 \texttt{sto\_32k} & \textbf{-2.74} & \textbf{-1.07} &  \textbf{3.75} &  \textbf{0.26} &  9.57 \\
\midrule
 \texttt{mp3\_64k} & \textbf{-2.64} & \textbf{-0.86} &  4.89 &  \textbf{0.07} & \textbf{17.85} \\
 \texttt{det\_64k} & -2.95 & -1.40 &  \textbf{3.54} &  0.16 & 12.13 \\
 \texttt{sto\_64k} & -2.74 & -1.02 &  3.59 &  0.17 & 11.51 \\
\bottomrule
\end{tabular}
\end{center}
\caption{Results of objective metrics for stochastic (\texttt{sto}), deterministic (\texttt{det}) models and MP3 baselines (\texttt{mp3}), for different compression rates ($16$kbit/s, $32$kbit/s, $64$kbit/s). Higher values  are better for ODG, DI and SNR; lower values are better for LSD and MSE.}
\label{tab:res_obj}
\end{table}

We test the method for three different MP3 compression rates ($16$kbit/s, $32$kbit/s and $64$kbit/s) as input to the generator.
Moreover, as stated above, we assume that there are multiple valid solutions for an MP3 to be restored with very high compression rates.
This would also mean that when using a stochastic generator, some of all possible samples should be closer to the original than when only using a deterministic generator.
In order to test this hypothesis, for each compression rate, we train a stochastic generator (with noise input $z$) and a deterministic generator (without noise input).
Then, for any input $y$ taken from the test set, we sample $20$ times with the corresponding generator using ${z}_i \sim \mathcal{N}(0,\mathbf{I})$, and for each objective metric, we take the best value of that set.
Note that all objective metrics are computed by comparing the restored data with the original versions. Therefore, when picking samples to optimize a specific metric, we do not pick the sample with the best ``quality'', but rather the restoration that best approximates the original.

Table \ref{tab:res_obj} and Figure \ref{fig:box_plots} show the results (i.e., the comparison to the high-quality data) for the stochastic and the deterministic models, and the respective MP3 baselines. For high compression rates (i.e., $16$kbit/s and $32$kbit/s), the best reconstructions of the stochastic models generally perform better than the baseline MP3s in most metrics and improve over the outputs of the deterministic models. This indicates that the facilitation of a stochastic generator is actually useful for restoration tasks. For some metrics (except LSD), the deterministic models perform on a par with the MP3 baselines. That is reasonable, as there are many different ways to restore the original version, and it is unlikely that a deterministic model outputs a close approximation. In Figure \ref{fig:box_plots} the strong violin-shaped forms in the figures indicate that the restorations form two groups in the ODG and DI metrics. From visual inspection of the respective data, it becomes clear that those excerpts in the lower (worse) groups are such without percussion instruments, indicating that the models cannot add meaningful high-frequency content for, e.g., singing voice or tonal instruments. The SNR is always worse for the restorations (compared to the MP3 baselines), which shows that the phase information is not faithfully regenerated. Given the high variety of possible phase information in the high frequency range, particularly for percussive sounds, this is not surprising, but also does not hamper the perceived audio quality.

For the $64$kbit/s MP3s, we see that the reconstructions are worse than the MP3 itself, except in the LSD metric. Note that $64$kbit/s mono MP3s are already close to the original. The fact that the generator performs worse on these data indicates that in addition to adding high frequency content (which is mostly advantageous, as can be seen in the LSD results), it also introduces some undesirable artifacts in the reconstruction of the MP3 information.

\subsubsection{Frequency Profiles}
In order to test the influence of the input noise $z$ onto the generator output, we input random MP3 examples and restore them while keeping the noise input fixed. Then, we calculate the frequency profiles of the resulting outputs by taking the mean over the time dimension. Figure \ref{fig:freq_profiles} shows examples of this experiment, which makes it clear that a specific $z$ causes a characteristic frequency profile consistently over different examples. This is advantageous when $z$ is chosen manually to control the restoration of an entire song, where a consistent characteristic is desired throughout the whole song.

\subsection{Subjective Evaluation}\label{sec:eval_subj}
In this section, we describe our own assessment when listening to the restored audio excerpts (Section \ref{sec:subj_assessment}), and then we provide results of the Mean Opinion Score (MOS), where we evaluate the restorations in a listening test with expert listeners.

\subsubsection{Informal Listening}\label{sec:subj_assessment}
For sound examples of the proposed method, please refer to the accompanying website.\footnote{\url{https://sonycslparis.github.io/restoration_mdpi_suppl_mat/}} When listening to the restored audio excerpts compared to the MP3 versions, the overall impression is a richer, higher bandwidth sound that could be described as ``opening up''. Also, we notice that the model is able to remove some MP3 artifacts, particularly \emph{swirlies}, as described in the introduction (see also \cite{sos}). It is clearly audible that the model adds frequency content which got lost in the MP3 compression. When comparing the restorations directly to the high-quality versions, it is noticeable that the level of detail in the high frequencies is considerably lower in the restorations. When inspecting the restorations closer, we can hear that for specific sound events, the model performs particularly well (i.e., adds convincing high-frequency content and removes specific compression artifacts), other sources do not undergo a considerable improvement, and some events tend to cause undesired, audible artifacts.

Among the sound events which are generally improved very well are percussive elements like snare, crash, hi-hat, and cymbal sounds, but also other onsets with steep transients and non-harmonic high-frequency content, like the strumming of acoustic guitars or sibilants or plosives ('s' and 't') in a singing voice. Also, sustained electric guitars undergo considerable improvement. Note that all these sound types do not possess harmonics but instead require the addition of high-frequency noise in the restoration process. Considering the nature of percussive sounds and the wide variety of sources in the training data, this is a reasonable outcome. On the one hand, percussive sounds dominate other sources in the higher frequency range, which constitutes the main difference between MP3 and high-quality versions of the audio excerpts. On the other hand, harmonic sources are extremely varied, and their harmonics are of different characteristics. In addition, harmonics are rarely found above 10kHz, which is the range in which the critic can best determine the difference between MP3 and high-quality audio signals.

Sometimes, the generator adds undesired, sustained noise, mainly when the audio input is very compressed or when there are rather loud, single tonal instruments or singing voice.  Other undesired artifacts added by the generator are mainly ``phantom percussions'', like hi-hats that do not have meaningful rhythmic positions, triggered by events in the MP3 input that get confused with percussive sources. Also, the generator sometimes overemphasizes 's' or 't' phonemes of a singing voice. However, in some cases percussive sounds not present in the original audio signals are added, which are rhythmically meaningful. In general, the overall characteristics of the percussion instruments are often different in the restorations compared to the high-quality versions. This is reasonable, as the lower frequencies present in the MP3 do not provide information about their characteristic in the higher frequency range, wherefore the characteristic needs to be regenerated by the model (dependent on the input noise $z$).

\subsubsection{Formal Listening}\label{sec:subj_eval}

\begin{table}[t]
    \centering
    \begin{tabular}{l r r}

        \toprule
          & mean & std \\
        \midrule
         original & 2.81 & 0.94 \\
         \midrule
         \texttt{mp3\_16k} & 0.74 & 0.79 \\
         \texttt{det\_16k} & 1.33 & 0.82\\
         \texttt{sto\_16k} & \textbf{1.40} & 0.89\\
         \midrule
         \texttt{mp3\_32k} & 0.80 & 0.71 \\
         \texttt{det\_32k} & \textbf{1.43} & 0.84\\
         \texttt{sto\_32k} & 1.28 & 0.82 \\
         \midrule
         \texttt{mp3\_64k} & \textbf{2.92} & 0.95\\
         \texttt{det\_64k} &  2.49 & 0.86\\
         \texttt{sto\_64k} & 2.65 & 0.74 \\
         \bottomrule
    \end{tabular}
    \caption{Mean Opinion Score (MOS) of absolute ratings for different compression rates. We compare the stochastic (\emph{sto}) versions against the deterministic baselines (\emph{det}), the MP3-encoded lower anchors (\emph{mp3}) and the \emph{original} high-quality audio excerpts.}

    \label{tab:subj_eval}
\end{table}

Table \ref{tab:subj_eval} shows the results of the listening test (i.e., MOS ratings). Overall, the \emph{original} and the $64$kbit/s MP3s (\texttt{mp3\_64k}) obtain the highest ratings and the restored $64$kbit/s MP3s (\texttt{det\_64k} and \texttt{sto\_64k}) perform slightly worse. The ratings for the restored $16$kbit/s and $32$kbit/s (\texttt{det\_16k}, \texttt{sto\_16k}, \texttt{det\_32k} and \texttt{sto\_32k}) are considerably better than the MP3 versions (\texttt{mp3\_16k} and \texttt{mp3\_32k}). This shows that the proposed restoration process indeed results in better perceived audio quality. However, the random samples from the stochastic generators are not assessed better than the outputs of the deterministic generators (the differences are not significant, as detailed below). We note that for the high compression rates, we reach only about half the average rating of the high-quality versions (but about double the rating of the MP3 versions). While overall, a restored MP3 version possesses a broader frequency range, weak ratings may result from off-putting artifacts, like the above-mentioned ``phantom percussions''. In $8$-second-long excerpts, only one irritating artifact can already lead to a relatively weak rating for the whole example.

As the variance of the ratings is rather high, we also compute t-tests for statistical significance comparing responses to the different stimuli. We obtain p-values $< 0.05$ ($<10^{-5}$) when comparing \texttt{det} and \texttt{sto} to \texttt{mp3} for compression rates below $64$kbit/s. Conversely, we observe no statistically significant differences between ratings of \texttt{det} and \texttt{sto} for all compression rates (p-values $>0.15$). Responses to \emph{original} and \texttt{mp3\_64k} also show no statistically significant differences (p-value $= 0.49$). We also observe no statistical significance between responses to \texttt{mp3\_64k} and \texttt{det\_64k} (p-value $ = 0.06$), whereas there is a significant difference between ratings of \texttt{sto\_64k} and \texttt{mp3\_64k} (p-value $= 0.04$).

\section{Conclusion and Future Work}\label{sec:conclusion}
\end{paracol}
\begin{figure}
\begin{tabular}{cc}
\includegraphics[trim=10 5 40 30,clip, width=0.42\textwidth]{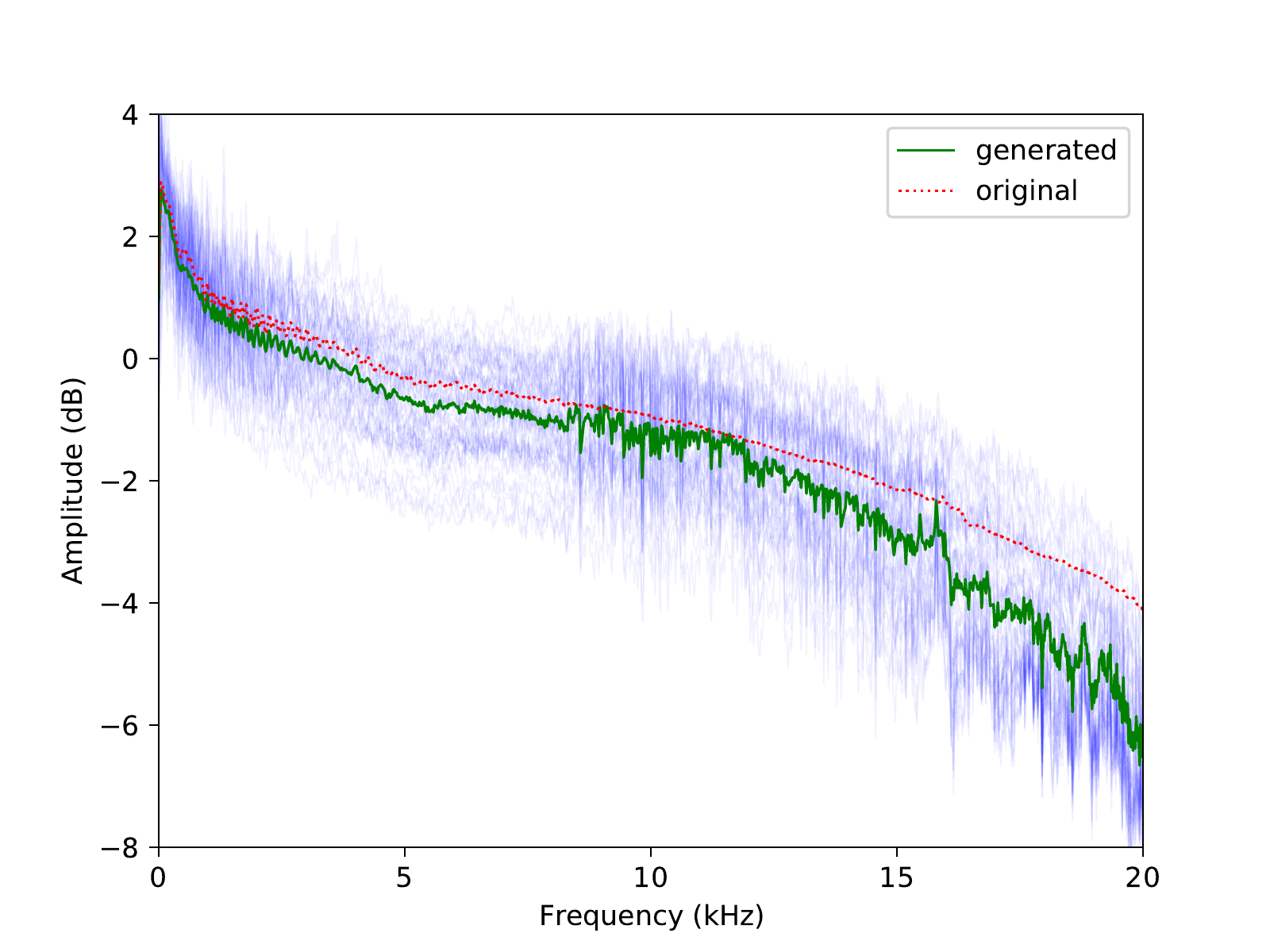} & 
\includegraphics[trim=10 5 40 30,clip, width=0.42\textwidth]{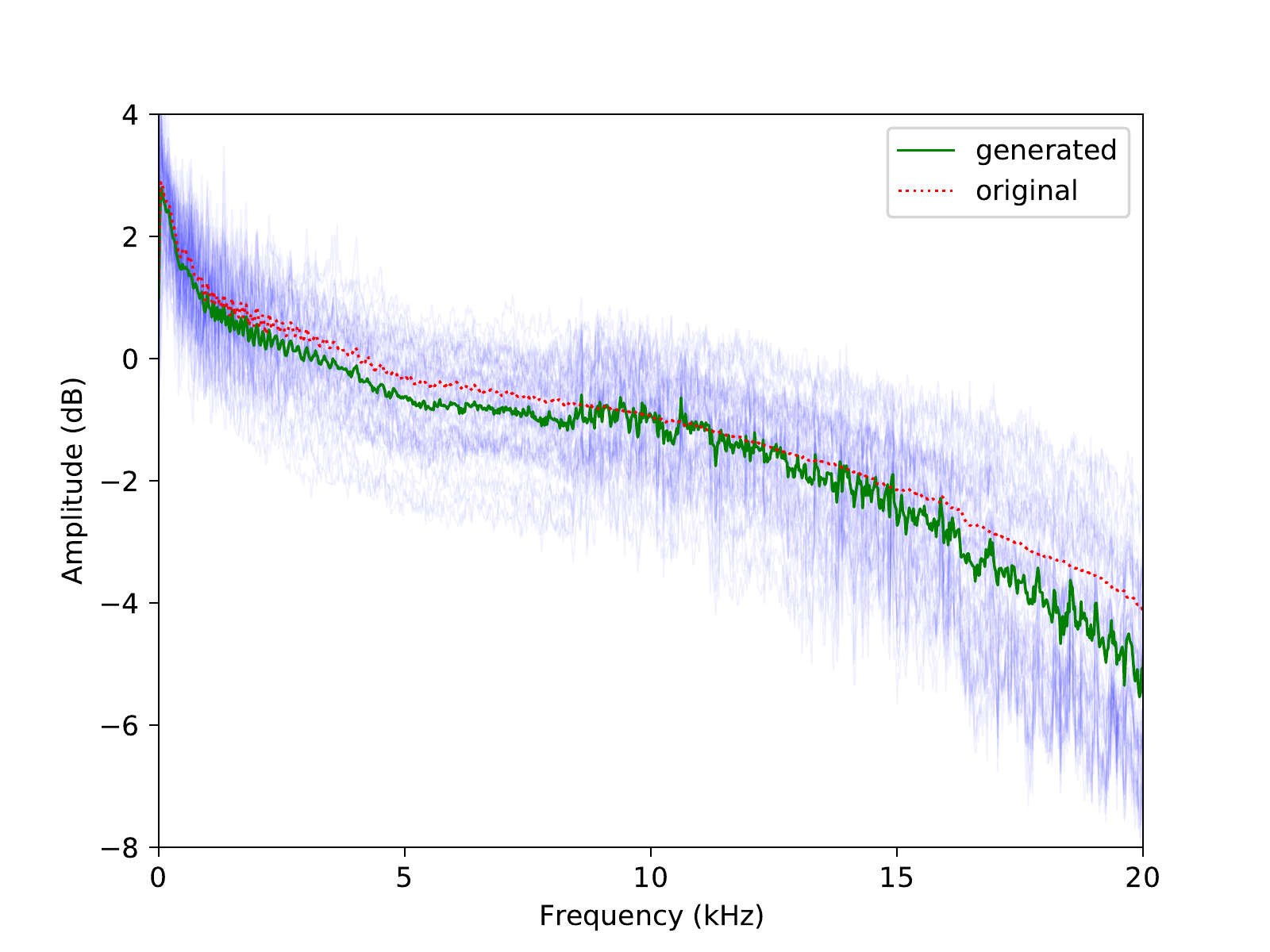} \\
\includegraphics[trim=10 5 40 30,clip, width=0.42\textwidth]{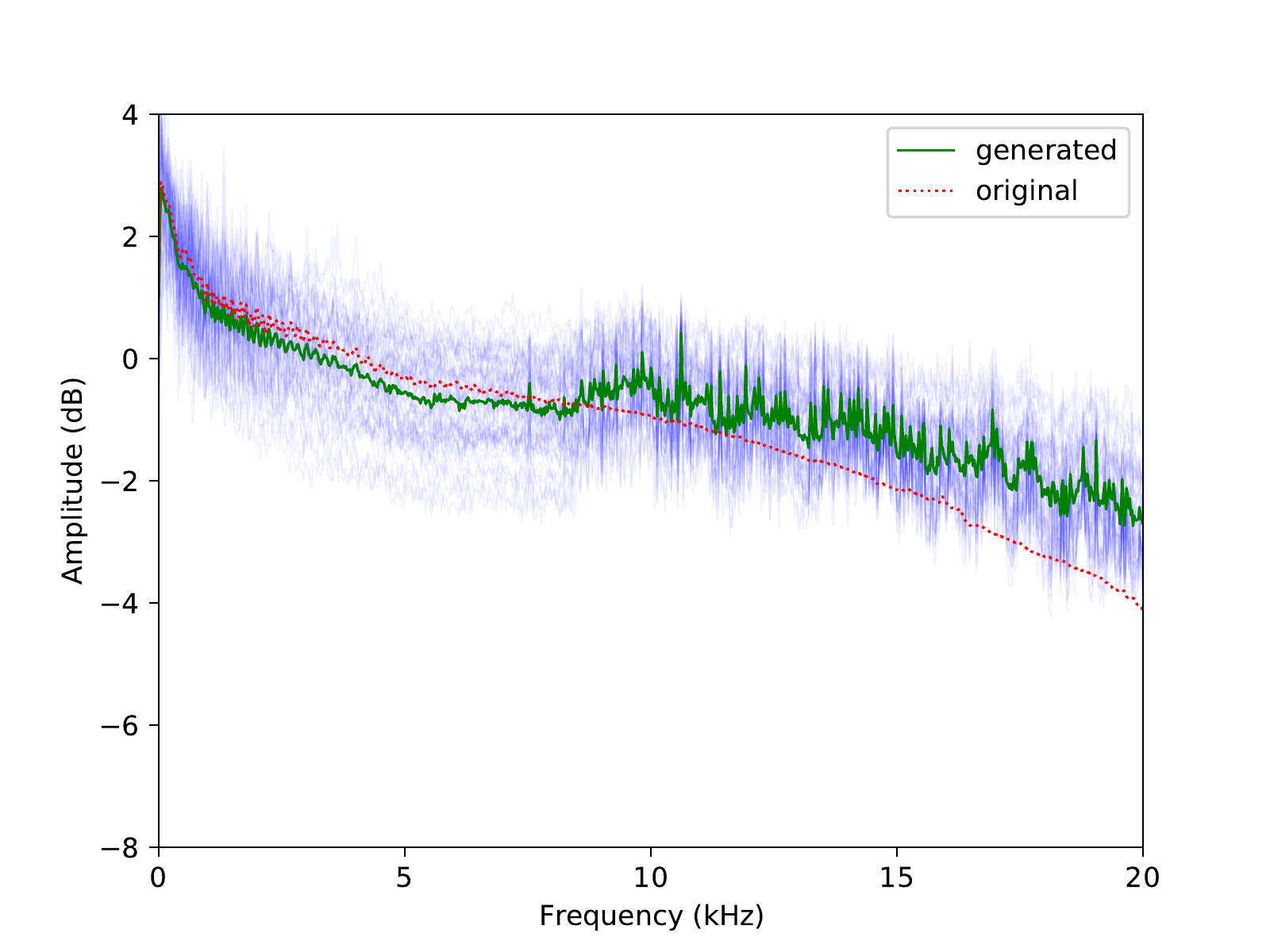} & 
\includegraphics[trim=10 5 40 30,clip, width=0.42\textwidth]{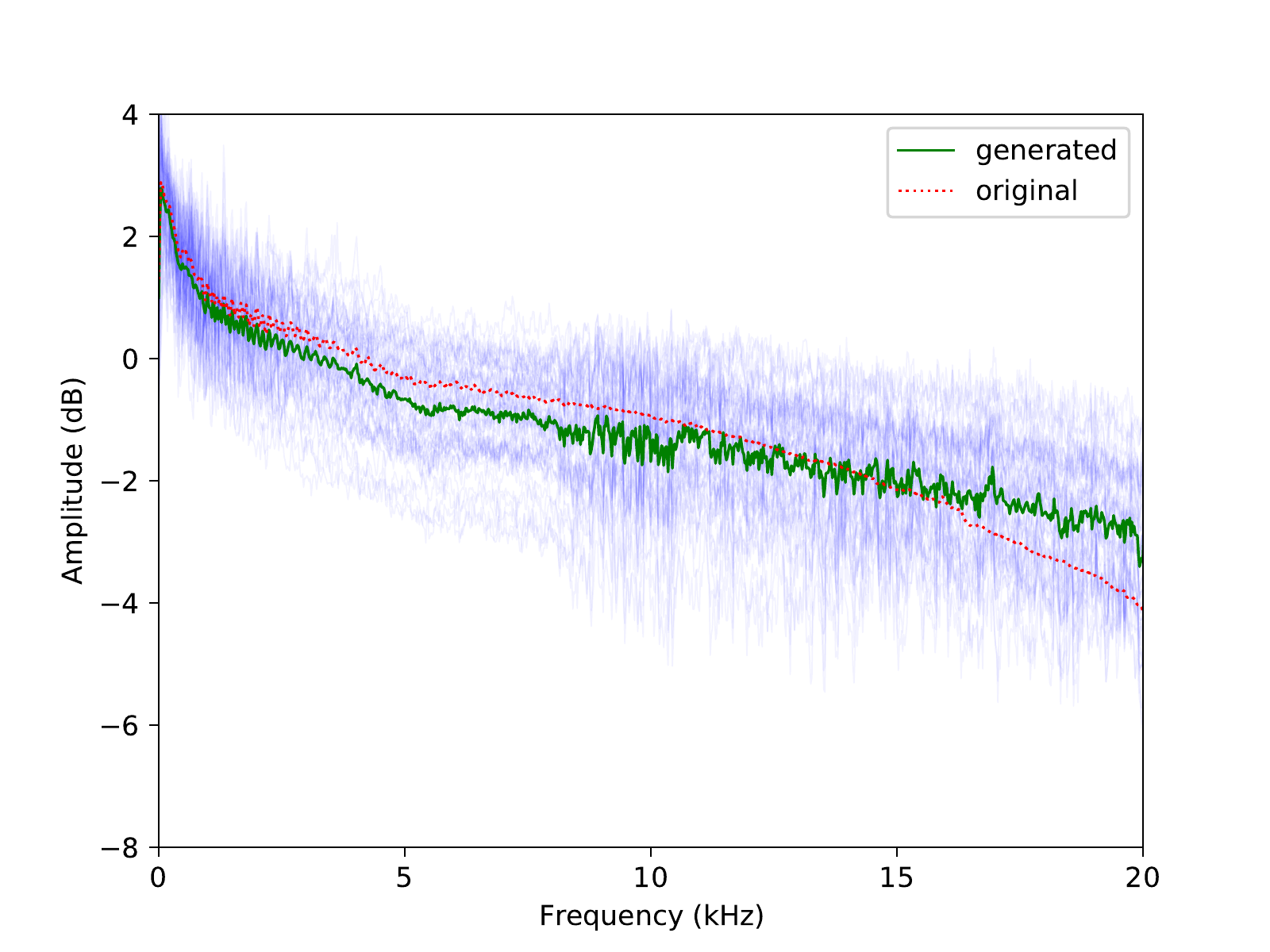} \\
\includegraphics[trim=10 5 40 30,clip, width=0.42\textwidth]{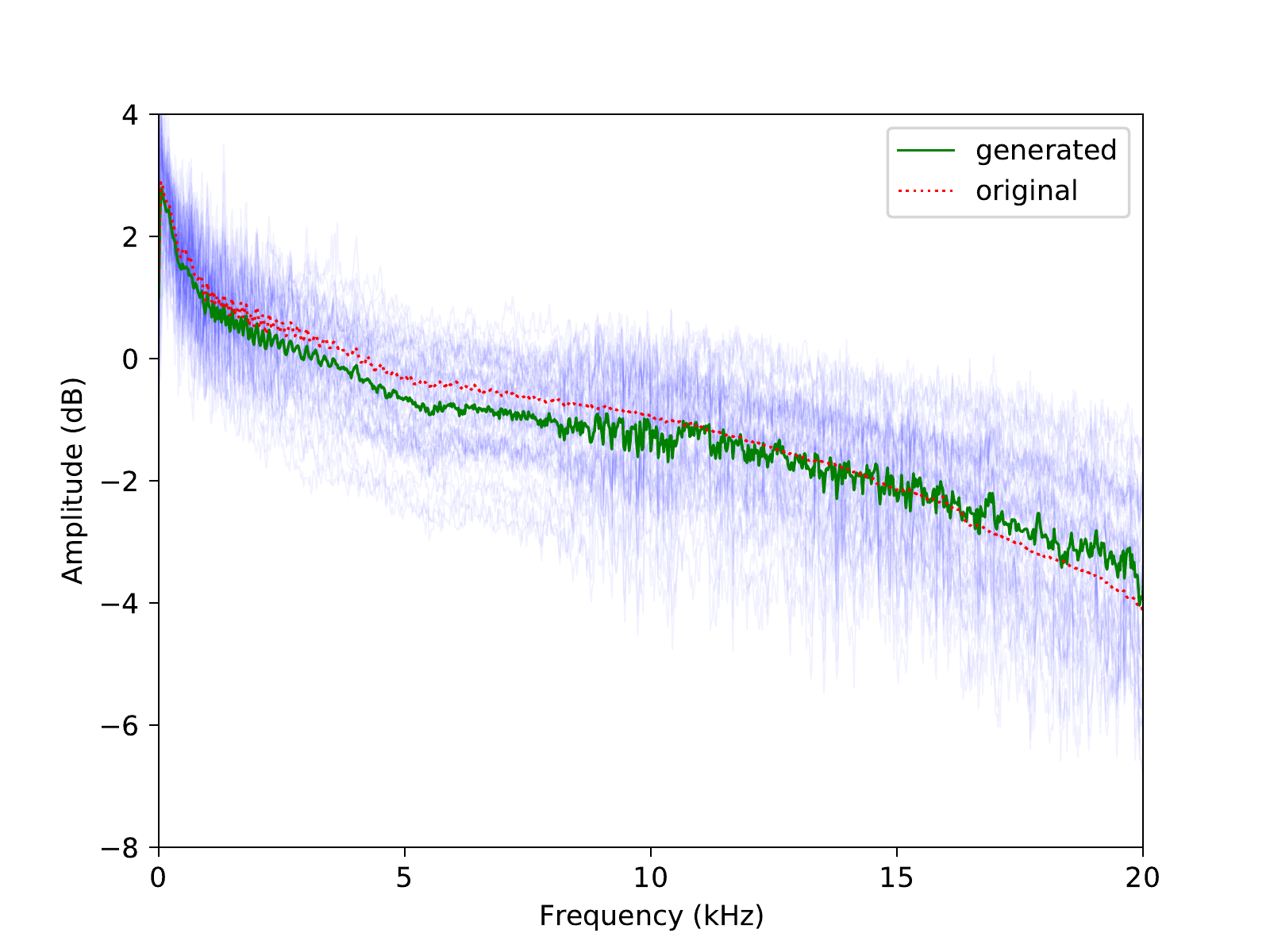} & 
\includegraphics[trim=10 5 40 30,clip, width=0.42\textwidth]{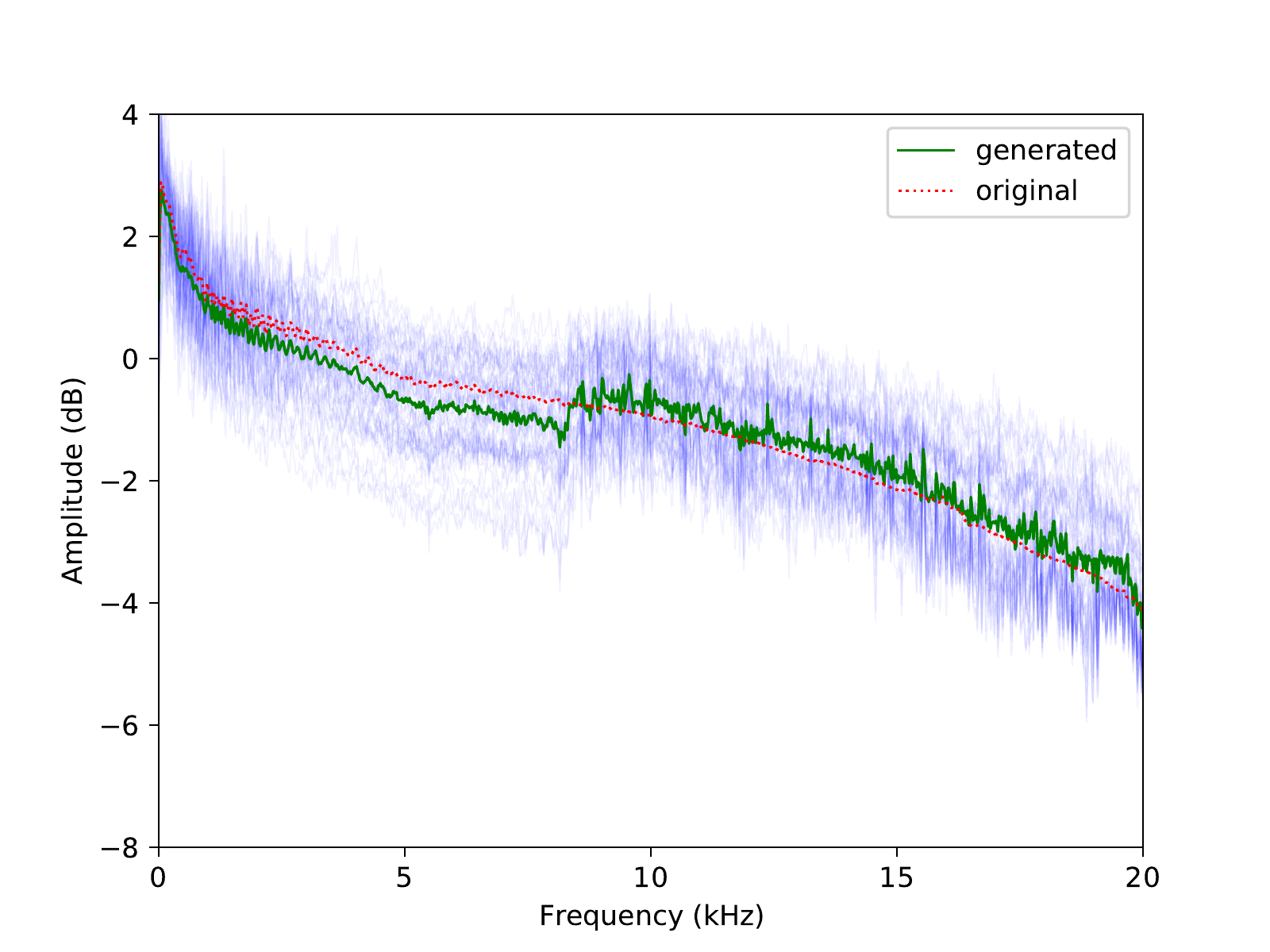} \\
\includegraphics[trim=10 5 40 30,clip, width=0.42\textwidth]{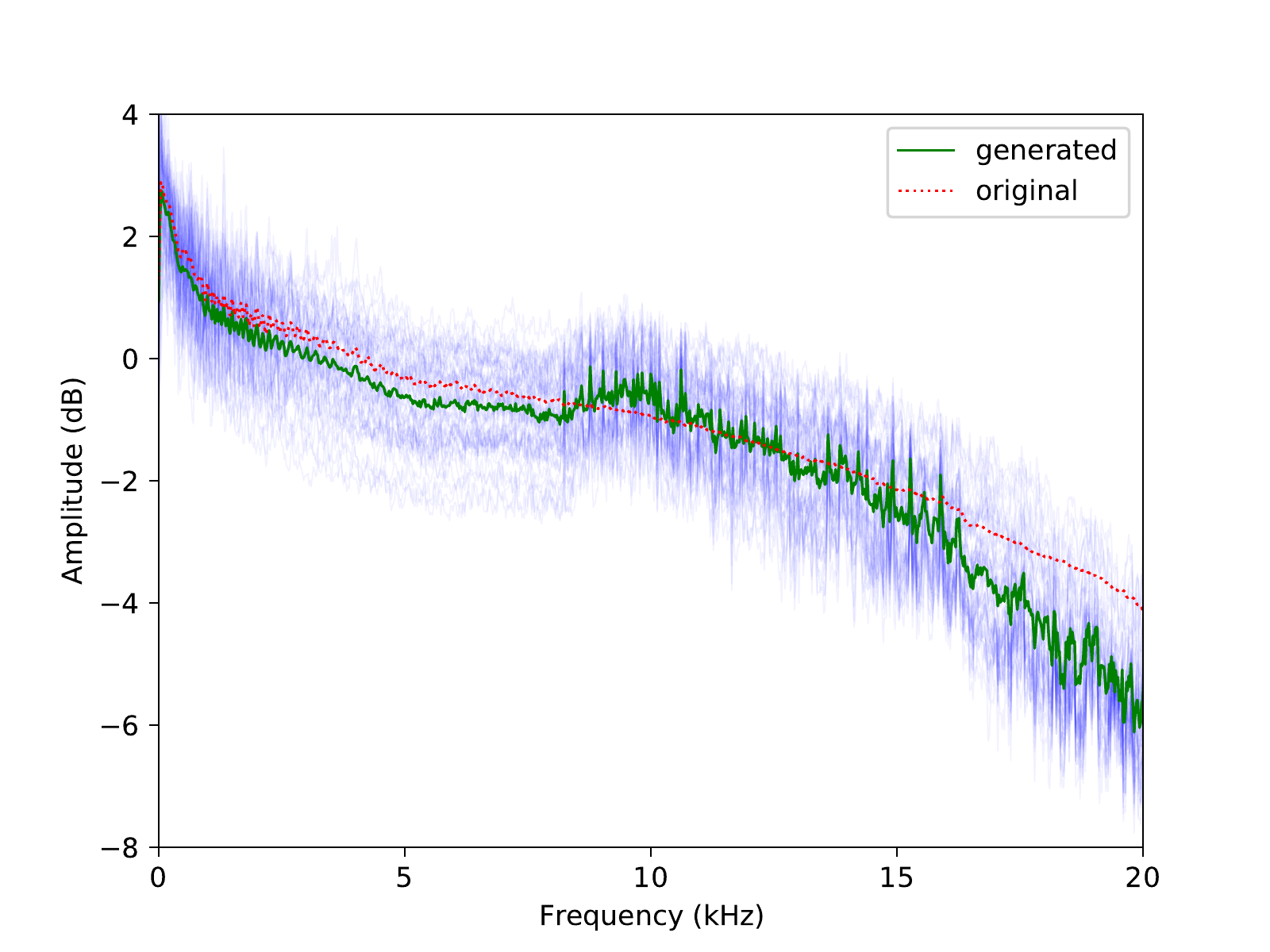} & 
\includegraphics[trim=10 5 40 30,clip, width=0.42\textwidth]{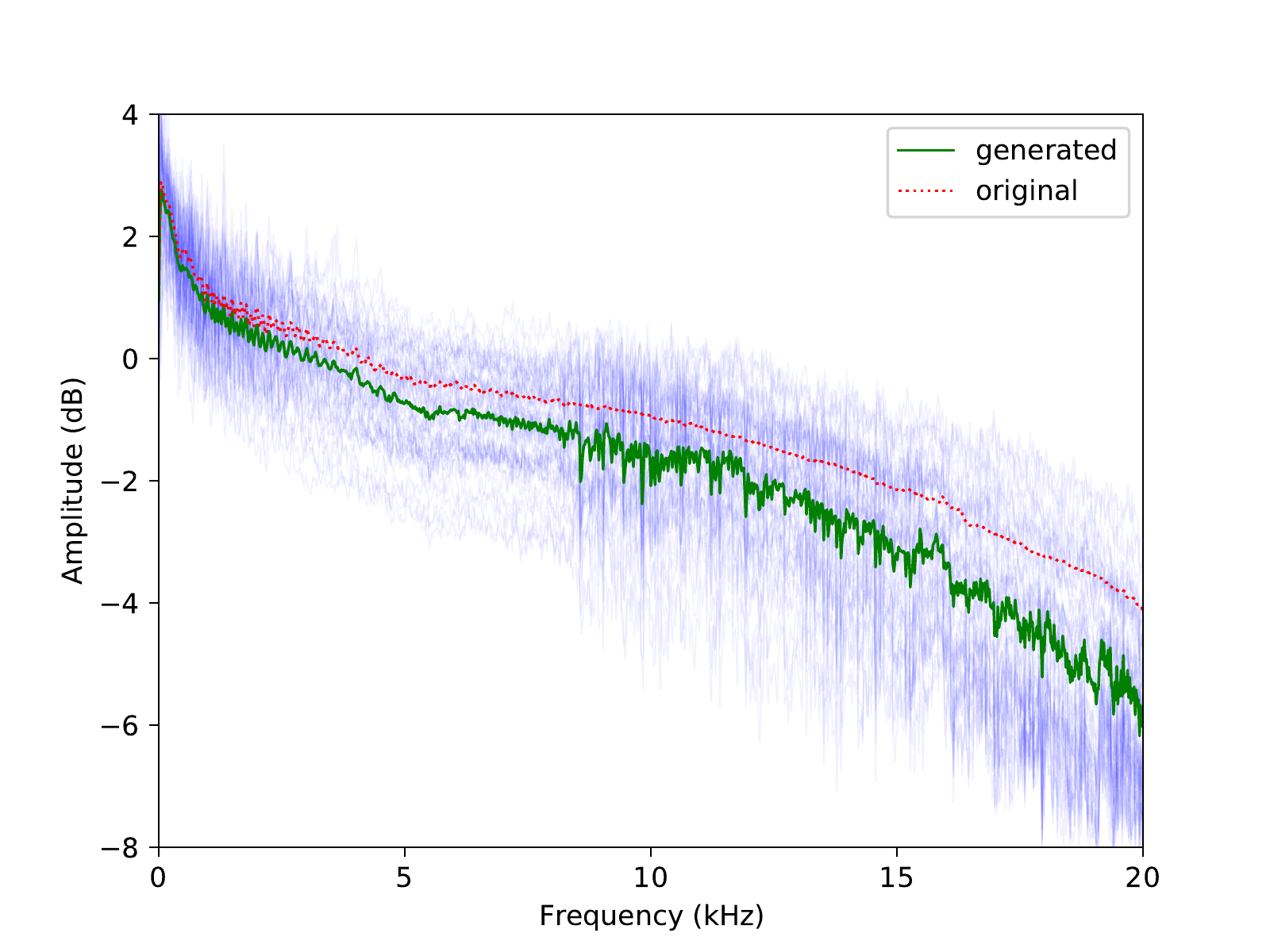} \\
\end{tabular}
\caption{Frequency profiles of $50$ random $4$-second-long excerpts from the test set (in $32$kbit/s) for different random input noise vectors ${z}$. The blue lines show the profiles of the individual samples, the green line shows the mean profile of the excerpts, the dotted red line shows the mean of the high-quality excerpts for comparison. It becomes clear that $z$ is strongly correlated with the energy in the upper bands and that a specific $z$ yields a consistent overall characteristic.}
\label{fig:freq_profiles}
\end{figure}
\begin{paracol}{2}
\switchcolumn
We presented a Generative Adversarial Network (GAN) architecture for stochastic restoration of high-quality musical audio signals from highly compressed MP3 versions. We tested 1) if the output of the proposed model improves the quality of the MP3 inputs, 2) if a stochastic generator improves (i.e., can generate samples closer to the original) over a deterministic generator, and 3) if the output of the stochastic variants are generally of higher quality than deterministic baseline models.

Results show that the restorations of the highly compressed MP3 versions (16kbit/s and 32kbit/s) are generally better than the MP3 versions themselves, which is reflected in a thorough objective evaluation, and confirmed in perceptual tests by human experts. We also tested weaker compression rates (64kbit/s mono), where we found that the proposed architecture results in slightly worse results than the MP3 baseline.
We could also show in the objective metrics that a stochastic generator can indeed output samples that are closer to the original than when using a deterministic generator. However, the perceptual tests indicate that when drawing random samples from the stochastic generator, the results are not assessed significantly better than the results of the deterministic generator.

Due to the wide variety of popular music, the task of generating missing content is very challenging.  However, the proposed models succeeded in adding high-frequency content for particular sources resulting in an overall improved perceived quality of the music. Examples for sources where the model clearly learned to generate meaningful high-frequency content are percussive elements (i.e., snare, crash, hi-hat and cymbal sounds), sibilants or plosives ('s' and 't') in singing voice, strummed acoustic guitars and (sustained) electric guitars.

We expect future improvements when limiting the style of the training data to particular genres or time periods of production. Also, as we use the complex spectrum directly, the adaption to Complex Networks \cite{trabelsi2018deep} could improve the results further. In order to tackle the problem of ``phantom percussions'' (as described in Section \ref{sec:subj_assessment}), a beat detection algorithm could provide additional information to the generator so that it is better informed about the rhythmic structure of the input. For improvement in learning to restore the harmonics of tonal sources, other representations (e.g., Magnitude and Instantaneous Frequencies (Mag-IF) \cite{DBLP:conf/iclr/EngelACGDR19}) or a different scaling (e.g., Mel-scaled spectrograms) could be tested for the input and output of the generator.

\section{Author biography}

Stefan Lattner is an associate researcher at Sony CSL Paris. He earned his Ph.D. degree in 2019 at the JKU Linz, at the CP Institute, under the supervision of Gerhard Widmer. He received his MSc. degree (in Informatics) in 2013 at the JKU Linz, and his BSc. degree (2009) in Mediatechnology and -design at the University of Applied Sciences in Hagenberg, Upper Austria.
From 2013 to 2018, Lattner worked at the Austrian Research Institute for Artificial Intelligence in Vienna, and between 2009 and 2014, he was lead developer and project manager at Re-Compose.
Lattner is concerned with musical structure learning, invariance learning, music and audio generation, as well as meta-learning and information theory in music cognition. He received the Best Paper Award for the paper ``Learning Complex Basis Functions for Invariant Representations of Audio'' at the ISMIR conference, Delft, 2019.

Javier Nistal is an assistant researcher at Sony CSL Paris and Ph.D. candidate at Telecom IP Paris under a Marie Curie fellowship. His current research is centered around controllable audio synthesis with Generative Adversarial Networks. Before starting his Ph.D., Nistal was a Machine Learning Researcher at MIDAS, a mixing console manufacturing company, which is part of Music Tribe. During his time at MIDAS, Nistal worked on instrument recognition and spill detection for the MIDAS Heritage D, the first mixing console which integrates Machine Learning. Before joining Music Tribe, Nistal interned in Jukedeck and SoundCloud as an MIR researcher. Nistal earned an MSc in Sound and Music Computing from the Music Technology Group (MTG) at Pompeu Fabra University and a BSc degree in Sound and Image Engineering from the Polytechnic University of Madrid.

\authorcontributions{Conceptualization, S.L.; methodology, S.L.; implementation,
S.L., J.N; obj. validation, S.L.; subj. validation, J.N., S.L.; investigation, S.L.; data curation,
S.L.; writing—original draft preparation, S.L., J.N.; writing—review and editing, S.L.;
visualization, S.L.; supervision, S.L.; project administration, S.L.; All authors have read and agreed to the published version of the manuscript.}

\funding{This research has received founding from the European Union’s Horizon 2020 research and innovation programme under the Marie Skłodowska-Curie grant agreement No. 765068}


\informedconsent{Informed consent was obtained from all subjects involved in the study. The listening study was performed voluntary, anonymous, and no unsettling material was presented to the subjects.}

\dataavailability{Generated audio examples produced in this study can be found at \url{https://sonycslparis.github.io/restoration_mdpi_suppl_mat/}.} 

\conflictsofinterest{The authors declare no conflict of interest.} 

\end{paracol}

\reftitle{References}




\bibliography{references.bib}

\begin{thebibliography}{999}

\bibitem[Brandenburg and Stoll(1994)]{mpeg-1}
Brandenburg, K.; Stoll, G.
\newblock ISO/MPEG-1 audio: A generic standard for coding of high-quality
  digital audio.
\newblock {\em Journal of the Audio Engineering Society} {\bf 1994}, {\em
  42},~780--792.

\bibitem[Brandenburg(1999)]{Karlheinz}
Brandenburg, K.
\newblock MP3 and AAC explained.
\newblock  Audio Engineering Society Conference: 17th International Conference:
  High-Quality Audio Coding. Audio Engineering Society,  1999.

\bibitem[Musmann(2006)]{DBLP:journals/tce/Musmann06}
Musmann, H.G.
\newblock Genesis of the {MP3} audio coding standard.
\newblock {\em {IEEE} Trans. Consumer Electron.} {\bf 2006}, {\em
  52},~1043--1049.
\newblock
  doi:{\changeurlcolor{black}\href{https://doi.org/10.1109/TCE.2006.1706505}{\detokenize{10.1109/TCE.2006.1706505}}}.

\bibitem[Corbett(2012)]{sos}
Corbett, I.
\newblock What data compression does to your music.
\newblock
  https://www.soundonsound.com/techniques/what-data-compression-does-your-music,
   2012.
\newblock Accessed 31 May 2021.

\bibitem[Williamson and Wang(2017)]{DBLP:conf/icassp/WilliamsonW17}
Williamson, D.S.; Wang, D.
\newblock Speech dereverberation and denoising using complex ratio masks.
\newblock  2017 {IEEE} International Conference on Acoustics, Speech and Signal
  Processing, {ICASSP} 2017, New Orleans, LA, USA, March 5-9, 2017. {IEEE},
  2017, pp. 5590--5594.
\newblock
  doi:{\changeurlcolor{black}\href{https://doi.org/10.1109/ICASSP.2017.7953226}{\detokenize{10.1109/ICASSP.2017.7953226}}}.

\bibitem[Kumar \em{et~al.}(2020)Kumar, Kumar, Anand, Bengio, and
  Courville]{DBLP:journals/corr/abs-2010-11362}
Kumar, R.; Kumar, K.; Anand, V.; Bengio, Y.; Courville, A.C.
\newblock {NU-GAN:} High resolution neural upsampling with {GAN}.
\newblock {\em CoRR} {\bf 2020}, {\em abs/2010.11362},
  \href{http://xxx.lanl.gov/abs/2010.11362}{{\normalfont [2010.11362]}}.

\bibitem[Zhao \em{et~al.}(2019)Zhao, Liu, and
  Fingscheidt]{DBLP:journals/taslp/ZhaoLF19}
Zhao, Z.; Liu, H.; Fingscheidt, T.
\newblock Convolutional Neural Networks to Enhance Coded Speech.
\newblock {\em {IEEE} {ACM} Trans. Audio Speech Lang. Process.} {\bf 2019},
  {\em 27},~663--678.
\newblock
  doi:{\changeurlcolor{black}\href{https://doi.org/10.1109/TASLP.2018.2887337}{\detokenize{10.1109/TASLP.2018.2887337}}}.

\bibitem[Fisher and Scherlis(2016)]{Fisher2016WaveMedicCN}
Fisher, K.; Scherlis, A.
\newblock WaveMedic: Convolutional Neural Networks for Speech Audio
  Enhancement.
\newblock  2016.

\bibitem[Skoglund and Valin(2020)]{DBLP:conf/interspeech/SkoglundV20}
Skoglund, J.; Valin, J.
\newblock Improving Opus Low Bit Rate Quality with Neural Speech Synthesis.
\newblock  Interspeech 2020, 21st Annual Conference of the International Speech
  Communication Association, Virtual Event, Shanghai, China, 25-29 October
  2020; Meng, H.; Xu, B.; Zheng, T.F., Eds. {ISCA},  2020, pp. 2847--2851.
\newblock
  doi:{\changeurlcolor{black}\href{https://doi.org/10.21437/Interspeech.2020-2939}{\detokenize{10.21437/Interspeech.2020-2939}}}.

\bibitem[Biswas and Jia(2020)]{DBLP:conf/icassp/BiswasJ20}
Biswas, A.; Jia, D.
\newblock Audio Codec Enhancement with Generative Adversarial Networks.
\newblock  2020 {IEEE} International Conference on Acoustics, Speech and Signal
  Processing, {ICASSP} 2020, Barcelona, Spain, May 4-8, 2020. {IEEE},  2020,
  pp. 356--360.
\newblock
  doi:{\changeurlcolor{black}\href{https://doi.org/10.1109/ICASSP40776.2020.9053113}{\detokenize{10.1109/ICASSP40776.2020.9053113}}}.

\bibitem[Porov \em{et~al.}(2018)Porov, Oh, Choo, Sung, Jeong, Osipov, and
  Francois]{porov2018music}
Porov, A.; Oh, E.; Choo, K.; Sung, H.; Jeong, J.; Osipov, K.; Francois, H.
\newblock Music enhancement by a novel CNN architecture.
\newblock  Audio Engineering Society Convention 145. Audio Engineering Society,
   2018.

\bibitem[Park and Lee(2017)]{DBLP:conf/interspeech/ParkL17}
Park, S.R.; Lee, J.
\newblock A Fully Convolutional Neural Network for Speech Enhancement.
\newblock  Interspeech 2017, 18th Annual Conference of the International Speech
  Communication Association, Stockholm, Sweden, August 20-24, 2017; Lacerda,
  F., Ed. {ISCA},  2017, pp. 1993--1997.

\bibitem[Gupta \em{et~al.}(2019)Gupta, Shillingford, Assael, and
  Walters]{DBLP:conf/waspaa/GuptaSAW19}
Gupta, A.; Shillingford, B.; Assael, Y.M.; Walters, T.C.
\newblock Speech Bandwidth Extension with Wavenet.
\newblock  2019 {IEEE} Workshop on Applications of Signal Processing to Audio
  and Acoustics, {WASPAA} 2019, New Paltz, NY, USA, October 20-23, 2019.
  {IEEE},  2019, pp. 205--208.
\newblock
  doi:{\changeurlcolor{black}\href{https://doi.org/10.1109/WASPAA.2019.8937169}{\detokenize{10.1109/WASPAA.2019.8937169}}}.

\bibitem[Isik \em{et~al.}(2020)Isik, Giri, Phansalkar, Valin, Helwani, and
  Krishnaswamy]{DBLP:conf/interspeech/IsikGPVHK20}
Isik, U.; Giri, R.; Phansalkar, N.; Valin, J.; Helwani, K.; Krishnaswamy, A.
\newblock PoCoNet: Better Speech Enhancement with Frequency-Positional
  Embeddings, Semi-Supervised Conversational Data, and Biased Loss.
\newblock  Interspeech 2020, 21st Annual Conference of the International Speech
  Communication Association, Virtual Event, Shanghai, China, 25-29 October
  2020; Meng, H.; Xu, B.; Zheng, T.F., Eds. {ISCA},  2020, pp. 2487--2491.
\newblock
  doi:{\changeurlcolor{black}\href{https://doi.org/10.21437/Interspeech.2020-3027}{\detokenize{10.21437/Interspeech.2020-3027}}}.

\bibitem[Hu \em{et~al.}(2020)Hu, Liu, Lv, Xing, Zhang, Fu, Wu, Zhang, and
  Xie]{DBLP:conf/interspeech/HuLLXZFWZX20}
Hu, Y.; Liu, Y.; Lv, S.; Xing, M.; Zhang, S.; Fu, Y.; Wu, J.; Zhang, B.; Xie,
  L.
\newblock {DCCRN:} Deep Complex Convolution Recurrent Network for Phase-Aware
  Speech Enhancement.
\newblock  Interspeech 2020, 21st Annual Conference of the International Speech
  Communication Association, Virtual Event, Shanghai, China, 25-29 October
  2020; Meng, H.; Xu, B.; Zheng, T.F., Eds. {ISCA},  2020, pp. 2472--2476.
\newblock
  doi:{\changeurlcolor{black}\href{https://doi.org/10.21437/Interspeech.2020-2537}{\detokenize{10.21437/Interspeech.2020-2537}}}.

\bibitem[Kontio \em{et~al.}(2007)Kontio, Laaksonen, and
  Alku]{DBLP:journals/taslp/KontioLA07}
Kontio, J.; Laaksonen, L.; Alku, P.
\newblock Neural Network-Based Artificial Bandwidth Expansion of Speech.
\newblock {\em {IEEE} Trans. Speech Audio Process.} {\bf 2007}, {\em
  15},~873--881.
\newblock
  doi:{\changeurlcolor{black}\href{https://doi.org/10.1109/TASL.2006.885934}{\detokenize{10.1109/TASL.2006.885934}}}.

\bibitem[Li and Lee(2015)]{DBLP:conf/icassp/LiL15}
Li, K.; Lee, C.
\newblock A deep neural network approach to speech bandwidth expansion.
\newblock  2015 {IEEE} International Conference on Acoustics, Speech and Signal
  Processing, {ICASSP} 2015, South Brisbane, Queensland, Australia, April
  19-24, 2015. {IEEE},  2015, pp. 4395--4399.
\newblock
  doi:{\changeurlcolor{black}\href{https://doi.org/10.1109/ICASSP.2015.7178801}{\detokenize{10.1109/ICASSP.2015.7178801}}}.

\bibitem[Xu \em{et~al.}(2015)Xu, Du, Dai, and Lee]{DBLP:journals/taslp/XuDDL15}
Xu, Y.; Du, J.; Dai, L.; Lee, C.
\newblock A Regression Approach to Speech Enhancement Based on Deep Neural
  Networks.
\newblock {\em {IEEE} {ACM} Trans. Audio Speech Lang. Process.} {\bf 2015},
  {\em 23},~7--19.
\newblock
  doi:{\changeurlcolor{black}\href{https://doi.org/10.1109/TASLP.2014.2364452}{\detokenize{10.1109/TASLP.2014.2364452}}}.

\bibitem[Lagrange and Gontier(2020)]{DBLP:conf/icassp/LagrangeG20}
Lagrange, M.; Gontier, F.
\newblock Bandwidth Extension of Musical Audio Signals With No Side Information
  Using Dilated Convolutional Neural Networks.
\newblock  2020 {IEEE} International Conference on Acoustics, Speech and Signal
  Processing, {ICASSP} 2020, Barcelona, Spain, May 4-8, 2020. {IEEE},  2020,
  pp. 801--805.
\newblock
  doi:{\changeurlcolor{black}\href{https://doi.org/10.1109/ICASSP40776.2020.9054194}{\detokenize{10.1109/ICASSP40776.2020.9054194}}}.

\bibitem[Miron and Davies(2018)]{Miron2018HIGHFM}
Miron, M.; Davies, M.
\newblock High frequency magnitude spectrogram reconstruction for music
  mixtures using convolutional autoencoders.
\newblock  Proc. of the 21st Int. Conference on Digital Audio Effects
  (DAFx-18). IEEE,  2018, pp. 173--180.

\bibitem[Deng \em{et~al.}(2020)Deng, Schuller, Eyben, Schuller, Zhang,
  Francois, and Oh]{DBLP:journals/nca/DengSESZFO20}
Deng, J.; Schuller, B.W.; Eyben, F.; Schuller, D.; Zhang, Z.; Francois, H.; Oh,
  E.
\newblock Exploiting time-frequency patterns with LSTM-RNNs for low-bitrate
  audio restoration.
\newblock {\em Neural Comput. Appl.} {\bf 2020}, {\em 32},~1095--1107.
\newblock
  doi:{\changeurlcolor{black}\href{https://doi.org/10.1007/s00521-019-04158-0}{\detokenize{10.1007/s00521-019-04158-0}}}.

\bibitem[Maiti and Mandel(2019)]{DBLP:conf/waspaa/MaitiM19}
Maiti, S.; Mandel, M.I.
\newblock Parametric Resynthesis With Neural Vocoders.
\newblock  2019 {IEEE} Workshop on Applications of Signal Processing to Audio
  and Acoustics, {WASPAA} 2019, New Paltz, NY, USA, October 20-23, 2019.
  {IEEE},  2019, pp. 303--307.
\newblock
  doi:{\changeurlcolor{black}\href{https://doi.org/10.1109/WASPAA.2019.8937165}{\detokenize{10.1109/WASPAA.2019.8937165}}}.

\bibitem[Dhariwal \em{et~al.}(2020)Dhariwal, Jun, Payne, Kim, Radford, and
  Sutskever]{DBLP:journals/corr/abs-2005-00341_long}
Dhariwal, P.; Jun, H.; Payne, C.; Kim, J.W.; Radford, A.; Sutskever, I.
\newblock Jukebox: {A} Generative Model for Music.
\newblock {\em CoRR} {\bf 2020}, {\em abs/2005.00341},
  \href{http://xxx.lanl.gov/abs/2005.00341}{{\normalfont [2005.00341]}}.

\bibitem[Goodfellow \em{et~al.}(2014)Goodfellow, Pouget-Abadie, Mirza, Xu,
  Warde-Farley, Ozair, Courville, and Bengio]{goodfellow2014generative}
Goodfellow, I.J.; Pouget-Abadie, J.; Mirza, M.; Xu, B.; Warde-Farley, D.;
  Ozair, S.; Courville, A.C.; Bengio, Y.
\newblock Generative Adversarial Nets.
\newblock  NIPS,  2014.

\bibitem[Nistal \em{et~al.}(2020)Nistal, Lattner, and Richard]{nistal2020}
Nistal, J.; Lattner, S.; Richard, G.
\newblock DrumGAN: Synthesis of Drum Sounds With Timbral Feature Conditioning
  Using Generative Adversarial Networks.
\newblock  Proceedings of the 21st International Society for Music Information
  Retrieval, {ISMIR}, Montreal, Canada,  2020.

\bibitem[Larsen and Aarts(2005)]{ABWE}
Larsen, E.; Aarts, R.M.
\newblock {\em Audio bandwidth extension: application of psychoacoustics,
  signal processing and loudspeaker design}; John Wiley \& Sons,  2005.

\bibitem[Bansal \em{et~al.}(2005)Bansal, Raj, and
  Smaragdis]{DBLP:conf/interspeech/BansalRS05}
Bansal, D.; Raj, B.; Smaragdis, P.
\newblock Bandwidth expansion of narrowband speech using non-negative matrix
  factorization.
\newblock  {INTERSPEECH} 2005 - Eurospeech, 9th European Conference on Speech
  Communication and Technology, Lisbon, Portugal, September 4-8, 2005. {ISCA},
  2005, pp. 1505--1508.

\bibitem[Makhoul and Berouti(1979)]{DBLP:conf/icassp/MakhoulB79}
Makhoul, J.; Berouti, M.G.
\newblock High-frequency regeneration in speech coding systems.
\newblock  {IEEE} International Conference on Acoustics, Speech, and Signal
  Processing, {ICASSP} '79, Washington, D. C., USA, April 2-4, 1979. {IEEE},
  1979, pp. 428--431.
\newblock
  doi:{\changeurlcolor{black}\href{https://doi.org/10.1109/ICASSP.1979.1170672}{\detokenize{10.1109/ICASSP.1979.1170672}}}.

\bibitem[Dietz \em{et~al.}(2002)Dietz, Liljeryd, Kjorling, and
  Kunz]{dietz2002spectral}
Dietz, M.; Liljeryd, L.; Kjorling, K.; Kunz, O.
\newblock Spectral Band Replication, a novel approach in audio coding.
\newblock  Audio Engineering Society Convention 112. Audio Engineering Society,
   2002.

\bibitem[Mandel and Cho(2015)]{DBLP:conf/waspaa/MandelC15}
Mandel, M.I.; Cho, Y.S.
\newblock Audio super-resolution using concatenative resynthesis.
\newblock  2015 {IEEE} Workshop on Applications of Signal Processing to Audio
  and Acoustics, {WASPAA} 2015, New Paltz, NY, USA, October 18-21, 2015.
  {IEEE},  2015, pp. 1--5.
\newblock
  doi:{\changeurlcolor{black}\href{https://doi.org/10.1109/WASPAA.2015.7336890}{\detokenize{10.1109/WASPAA.2015.7336890}}}.

\bibitem[Dong \em{et~al.}(2015)Dong, Wang, and
  Chambers]{DBLP:conf/icdsp/DongWC15}
Dong, J.; Wang, W.; Chambers, J.A.
\newblock Audio super-resolution using analysis dictionary learning.
\newblock  2015 {IEEE} International Conference on Digital Signal Processing,
  {DSP} 2015, Singapore, July 21-24, 2015. {IEEE},  2015, pp. 604--608.
\newblock
  doi:{\changeurlcolor{black}\href{https://doi.org/10.1109/ICDSP.2015.7251945}{\detokenize{10.1109/ICDSP.2015.7251945}}}.

\bibitem[Dong \em{et~al.}(2016)Dong, Loy, He, and
  Tang]{DBLP:journals/pami/DongLHT16}
Dong, C.; Loy, C.C.; He, K.; Tang, X.
\newblock Image Super-Resolution Using Deep Convolutional Networks.
\newblock {\em {IEEE} Trans. Pattern Anal. Mach. Intell.} {\bf 2016}, {\em
  38},~295--307.
\newblock
  doi:{\changeurlcolor{black}\href{https://doi.org/10.1109/TPAMI.2015.2439281}{\detokenize{10.1109/TPAMI.2015.2439281}}}.

\bibitem[Isola \em{et~al.}(2017)Isola, Zhu, Zhou, and
  Efros]{DBLP:conf/cvpr/IsolaZZE17}
Isola, P.; Zhu, J.; Zhou, T.; Efros, A.A.
\newblock Image-to-Image Translation with Conditional Adversarial Networks.
\newblock  2017 {IEEE} Conference on Computer Vision and Pattern Recognition,
  {CVPR} 2017, Honolulu, HI, USA, July 21-26, 2017. {IEEE} Computer Society,
  2017, pp. 5967--5976.
\newblock
  doi:{\changeurlcolor{black}\href{https://doi.org/10.1109/CVPR.2017.632}{\detokenize{10.1109/CVPR.2017.632}}}.

\bibitem[Kuleshov \em{et~al.}(2017)Kuleshov, Enam, and
  Ermon]{DBLP:conf/iclr/KuleshovEE17}
Kuleshov, V.; Enam, S.Z.; Ermon, S.
\newblock Audio Super-Resolution using Neural Networks.
\newblock  5th International Conference on Learning Representations, {ICLR}
  2017, Toulon, France, April 24-26, 2017, Workshop Track Proceedings.
  OpenReview.net,  2017.

\bibitem[Lim \em{et~al.}(2018)Lim, Yeh, Xu, Do, and
  Hasegawa{-}Johnson]{DBLP:conf/icassp/LimYXDH18}
Lim, T.; Yeh, R.A.; Xu, Y.; Do, M.N.; Hasegawa{-}Johnson, M.
\newblock Time-Frequency Networks for Audio Super-Resolution.
\newblock  2018 {IEEE} International Conference on Acoustics, Speech and Signal
  Processing, {ICASSP} 2018, Calgary, AB, Canada, April 15-20, 2018. {IEEE},
  2018, pp. 646--650.
\newblock
  doi:{\changeurlcolor{black}\href{https://doi.org/10.1109/ICASSP.2018.8462049}{\detokenize{10.1109/ICASSP.2018.8462049}}}.

\bibitem[Ling \em{et~al.}(2018)Ling, Ai, Gu, and
  Dai]{DBLP:journals/taslp/LingAGD18}
Ling, Z.; Ai, Y.; Gu, Y.; Dai, L.
\newblock Waveform Modeling and Generation Using Hierarchical Recurrent Neural
  Networks for Speech Bandwidth Extension.
\newblock {\em {IEEE} {ACM} Trans. Audio Speech Lang. Process.} {\bf 2018},
  {\em 26},~883--894.
\newblock
  doi:{\changeurlcolor{black}\href{https://doi.org/10.1109/TASLP.2018.2798811}{\detokenize{10.1109/TASLP.2018.2798811}}}.

\bibitem[Loizou(2007)]{speech_enhancement}
Loizou, P.
\newblock {\em Speech Enhancement: Theory and Practice};  2007.
\newblock
  doi:{\changeurlcolor{black}\href{https://doi.org/10.1201/b14529}{\detokenize{10.1201/b14529}}}.

\bibitem[Ortega{-}Garcia and
  Gonzalez{-}Rodriguez(1996)]{DBLP:conf/interspeech/Ortega-GarciaG96}
Ortega{-}Garcia, J.; Gonzalez{-}Rodriguez, J.
\newblock Overview of speech enhancement techniques for automatic speaker
  recognition.
\newblock  The 4th International Conference on Spoken Language Processing,
  Philadelphia, PA, USA, October 3-6, 1996. {ISCA},  1996.

\bibitem[Seltzer \em{et~al.}(2013)Seltzer, Yu, and
  Wang]{DBLP:conf/icassp/SeltzerYW13}
Seltzer, M.L.; Yu, D.; Wang, Y.
\newblock An investigation of deep neural networks for noise robust speech
  recognition.
\newblock  {IEEE} International Conference on Acoustics, Speech and Signal
  Processing, {ICASSP} 2013, Vancouver, BC, Canada, May 26-31, 2013. {IEEE},
  2013, pp. 7398--7402.
\newblock
  doi:{\changeurlcolor{black}\href{https://doi.org/10.1109/ICASSP.2013.6639100}{\detokenize{10.1109/ICASSP.2013.6639100}}}.

\bibitem[Kolb{\ae}k \em{et~al.}(2016)Kolb{\ae}k, Tan, and
  Jensen]{DBLP:conf/slt/KolboekTJ16}
Kolb{\ae}k, M.; Tan, Z.; Jensen, J.
\newblock Speech enhancement using Long Short-Term Memory based recurrent
  Neural Networks for noise robust Speaker Verification.
\newblock  2016 {IEEE} Spoken Language Technology Workshop, {SLT} 2016, San
  Diego, CA, USA, December 13-16, 2016. {IEEE},  2016, pp. 305--311.
\newblock
  doi:{\changeurlcolor{black}\href{https://doi.org/10.1109/SLT.2016.7846281}{\detokenize{10.1109/SLT.2016.7846281}}}.

\bibitem[Yang and Fu(2005)]{cochlear}
Yang, L.P.; Fu, Q.J.
\newblock Spectral subtraction-based speech enhancement for cochlear implant
  patients in background noise (L).
\newblock {\em The Journal of the Acoustical Society of America} {\bf 2005},
  {\em 117},~1001--4.
\newblock
  doi:{\changeurlcolor{black}\href{https://doi.org/10.1121/1.1852873}{\detokenize{10.1121/1.1852873}}}.

\bibitem[Chen \em{et~al.}(2016)Chen, Wang, Yoho, Wang, and Healy]{Jitong}
Chen, J.; Wang, Y.; Yoho, S.; Wang, D.; Healy, E.
\newblock Large-scale training to increase speech intelligibility for
  hearing-impaired listeners in novel noises.
\newblock {\em The Journal of the Acoustical Society of America} {\bf 2016},
  {\em 139},~2604--2612.
\newblock
  doi:{\changeurlcolor{black}\href{https://doi.org/10.1121/1.4948445}{\detokenize{10.1121/1.4948445}}}.

\bibitem[{Boll}(1979)]{1163209}
{Boll}, S.
\newblock Suppression of acoustic noise in speech using spectral subtraction.
\newblock {\em IEEE Transactions on Acoustics, Speech, and Signal Processing}
  {\bf 1979}, {\em 27},~113--120.
\newblock
  doi:{\changeurlcolor{black}\href{https://doi.org/10.1109/TASSP.1979.1163209}{\detokenize{10.1109/TASSP.1979.1163209}}}.

\bibitem[{Jae Lim} and {Oppenheim}(1978)]{1163086}
{Jae Lim}.; {Oppenheim}, A.
\newblock All-pole modeling of degraded speech.
\newblock {\em IEEE Transactions on Acoustics, Speech, and Signal Processing}
  {\bf 1978}, {\em 26},~197--210.
\newblock
  doi:{\changeurlcolor{black}\href{https://doi.org/10.1109/TASSP.1978.1163086}{\detokenize{10.1109/TASSP.1978.1163086}}}.

\bibitem[{Ephraim}(1992)]{168664}
{Ephraim}, Y.
\newblock Statistical-model-based speech enhancement systems.
\newblock {\em Proceedings of the IEEE} {\bf 1992}, {\em 80},~1526--1555.
\newblock
  doi:{\changeurlcolor{black}\href{https://doi.org/10.1109/5.168664}{\detokenize{10.1109/5.168664}}}.

\bibitem[Dendrinos \em{et~al.}(1991)Dendrinos, Bakamidis, and
  Carayannis]{DBLP:journals/speech/DendrinosBC91}
Dendrinos, M.; Bakamidis, S.; Carayannis, G.
\newblock Speech enhancement from noise: {A} regenerative approach.
\newblock {\em Speech Commun.} {\bf 1991}, {\em 10},~45--57.
\newblock
  doi:{\changeurlcolor{black}\href{https://doi.org/10.1016/0167-6393(91)90027-Q}{\detokenize{10.1016/0167-6393(91)90027-Q}}}.

\bibitem[Williamson \em{et~al.}(2016)Williamson, Wang, and
  Wang]{DBLP:journals/taslp/WilliamsonWW16}
Williamson, D.S.; Wang, Y.; Wang, D.
\newblock Complex Ratio Masking for Monaural Speech Separation.
\newblock {\em {IEEE} {ACM} Trans. Audio Speech Lang. Process.} {\bf 2016},
  {\em 24},~483--492.
\newblock
  doi:{\changeurlcolor{black}\href{https://doi.org/10.1109/TASLP.2015.2512042}{\detokenize{10.1109/TASLP.2015.2512042}}}.

\bibitem[Erdogan \em{et~al.}(2015)Erdogan, Hershey, Watanabe, and
  Roux]{DBLP:conf/icassp/ErdoganHWR15}
Erdogan, H.; Hershey, J.R.; Watanabe, S.; Roux, J.L.
\newblock Phase-sensitive and recognition-boosted speech separation using deep
  recurrent neural networks.
\newblock  2015 {IEEE} International Conference on Acoustics, Speech and Signal
  Processing, {ICASSP} 2015, South Brisbane, Queensland, Australia, April
  19-24, 2015. {IEEE},  2015, pp. 708--712.
\newblock
  doi:{\changeurlcolor{black}\href{https://doi.org/10.1109/ICASSP.2015.7178061}{\detokenize{10.1109/ICASSP.2015.7178061}}}.

\bibitem[Pascual \em{et~al.}(2017)Pascual, Bonafonte, and
  Serr{\`{a}}]{DBLP:conf/interspeech/PascualBS17}
Pascual, S.; Bonafonte, A.; Serr{\`{a}}, J.
\newblock {SEGAN:} Speech Enhancement Generative Adversarial Network.
\newblock  Interspeech 2017, 18th Annual Conference of the International Speech
  Communication Association, Stockholm, Sweden, August 20-24, 2017; Lacerda,
  F., Ed. {ISCA},  2017, pp. 3642--3646.

\bibitem[Pascual \em{et~al.}(2019)Pascual, Serr{\`{a}}, and
  Bonafonte]{DBLP:conf/interspeech/PascualSB19}
Pascual, S.; Serr{\`{a}}, J.; Bonafonte, A.
\newblock Towards Generalized Speech Enhancement with Generative Adversarial
  Networks.
\newblock  Interspeech 2019, 20th Annual Conference of the International Speech
  Communication Association, Graz, Austria, 15-19 September 2019; Kubin, G.;
  Kacic, Z., Eds. {ISCA},  2019, pp. 1791--1795.
\newblock
  doi:{\changeurlcolor{black}\href{https://doi.org/10.21437/Interspeech.2019-2688}{\detokenize{10.21437/Interspeech.2019-2688}}}.

\bibitem[Li \em{et~al.}(2018)Li, Dai, Song, and
  McLoughlin]{DBLP:journals/cssp/LiDSM18}
Li, Z.; Dai, L.; Song, Y.; McLoughlin, I.V.
\newblock A Conditional Generative Model for Speech Enhancement.
\newblock {\em Circuits Syst. Signal Process.} {\bf 2018}, {\em
  37},~5005--5022.
\newblock
  doi:{\changeurlcolor{black}\href{https://doi.org/10.1007/s00034-018-0798-4}{\detokenize{10.1007/s00034-018-0798-4}}}.

\bibitem[Donahue \em{et~al.}(2018)Donahue, Li, and
  Prabhavalkar]{DBLP:conf/icassp/DonahueLP18}
Donahue, C.; Li, B.; Prabhavalkar, R.
\newblock Exploring Speech Enhancement with Generative Adversarial Networks for
  Robust Speech Recognition.
\newblock  2018 {IEEE} International Conference on Acoustics, Speech and Signal
  Processing, {ICASSP} 2018, Calgary, AB, Canada, April 15-20, 2018. {IEEE},
  2018, pp. 5024--5028.
\newblock
  doi:{\changeurlcolor{black}\href{https://doi.org/10.1109/ICASSP.2018.8462581}{\detokenize{10.1109/ICASSP.2018.8462581}}}.

\bibitem[Michelsanti and Tan(2017)]{DBLP:conf/interspeech/MichelsantiT17}
Michelsanti, D.; Tan, Z.
\newblock Conditional Generative Adversarial Networks for Speech Enhancement
  and Noise-Robust Speaker Verification.
\newblock  Interspeech 2017, 18th Annual Conference of the International Speech
  Communication Association, Stockholm, Sweden, August 20-24, 2017; Lacerda,
  F., Ed. {ISCA},  2017, pp. 2008--2012.

\bibitem[Fu \em{et~al.}(2019)Fu, Liao, Tsao, and Lin]{DBLP:conf/icml/FuLTL19}
Fu, S.; Liao, C.; Tsao, Y.; Lin, S.
\newblock MetricGAN: Generative Adversarial Networks based Black-box Metric
  Scores Optimization for Speech Enhancement.
\newblock  Proceedings of the 36th International Conference on Machine
  Learning, {ICML} 2019, 9-15 June 2019, Long Beach, California, {USA};
  Chaudhuri, K.; Salakhutdinov, R., Eds. {PMLR},  2019, Vol.~97, {\em
  Proceedings of Machine Learning Research}, pp. 2031--2041.

\bibitem[Phan \em{et~al.}(2020)Phan, McLoughlin, Pham, Ch{\'{e}}n, Koch, Vos,
  and Mertins]{DBLP:journals/spl/PhanMPCKVM20}
Phan, H.; McLoughlin, I.V.; Pham, L.D.; Ch{\'{e}}n, O.Y.; Koch, P.; Vos, M.D.;
  Mertins, A.
\newblock Improving GANs for Speech Enhancement.
\newblock {\em {IEEE} Signal Process. Lett.} {\bf 2020}, {\em 27},~1700--1704.
\newblock
  doi:{\changeurlcolor{black}\href{https://doi.org/10.1109/LSP.2020.3025020}{\detokenize{10.1109/LSP.2020.3025020}}}.

\bibitem[Germain \em{et~al.}(2019)Germain, Chen, and
  Koltun]{DBLP:conf/interspeech/GermainCK19}
Germain, F.G.; Chen, Q.; Koltun, V.
\newblock Speech Denoising with Deep Feature Losses.
\newblock  Interspeech 2019, 20th Annual Conference of the International Speech
  Communication Association, Graz, Austria, 15-19 September 2019; Kubin, G.;
  Kacic, Z., Eds. {ISCA},  2019, pp. 2723--2727.
\newblock
  doi:{\changeurlcolor{black}\href{https://doi.org/10.21437/Interspeech.2019-1924}{\detokenize{10.21437/Interspeech.2019-1924}}}.

\bibitem[Su \em{et~al.}(2020)Su, Jin, and
  Finkelstein]{DBLP:conf/interspeech/SuJF20}
Su, J.; Jin, Z.; Finkelstein, A.
\newblock HiFi-GAN: High-Fidelity Denoising and Dereverberation Based on Speech
  Deep Features in Adversarial Networks.
\newblock  Interspeech 2020, 21st Annual Conference of the International Speech
  Communication Association, Virtual Event, Shanghai, China, 25-29 October
  2020; Meng, H.; Xu, B.; Zheng, T.F., Eds. {ISCA},  2020, pp. 4506--4510.
\newblock
  doi:{\changeurlcolor{black}\href{https://doi.org/10.21437/Interspeech.2020-2143}{\detokenize{10.21437/Interspeech.2020-2143}}}.

\bibitem[Trabelsi \em{et~al.}(2018)Trabelsi, Bilaniuk, Zhang, Serdyuk,
  Subramanian, Santos, Mehri, Rostamzadeh, Bengio, and
  Pal]{DBLP:conf/iclr/TrabelsiBZSSSMR18}
Trabelsi, C.; Bilaniuk, O.; Zhang, Y.; Serdyuk, D.; Subramanian, S.; Santos,
  J.F.; Mehri, S.; Rostamzadeh, N.; Bengio, Y.; Pal, C.J.
\newblock Deep Complex Networks.
\newblock  6th International Conference on Learning Representations, {ICLR}
  2018, Vancouver, BC, Canada, April 30 - May 3, 2018, Conference Track
  Proceedings. OpenReview.net,  2018.

\bibitem[Zhao \em{et~al.}(2021)Zhao, Nguyen, and
  Ma]{DBLP:journals/corr/abs-2102-01993}
Zhao, S.; Nguyen, T.H.; Ma, B.
\newblock Monaural Speech Enhancement with Complex Convolutional Block
  Attention Module and Joint Time Frequency Losses.
\newblock {\em CoRR} {\bf 2021}, {\em abs/2102.01993},
  \href{http://xxx.lanl.gov/abs/2102.01993}{{\normalfont [2102.01993]}}.

\bibitem[Arjovsky \em{et~al.}(2017)Arjovsky, Chintala, and
  Bottou]{pmlr-v70-arjovsky17a}
Arjovsky, M.; Chintala, S.; Bottou, L.
\newblock {W}asserstein Generative Adversarial Networks.
\newblock  Proceedings of the 34th International Conference on Machine
  Learning; Precup, D.; Teh, Y.W., Eds. PMLR,  2017, Vol.~70, {\em Proceedings
  of Machine Learning Research}, pp. 214--223.

\bibitem[Long \em{et~al.}(2015)Long, Shelhamer, and
  Darrell]{DBLP:conf/cvpr/LongSD15}
Long, J.; Shelhamer, E.; Darrell, T.
\newblock Fully convolutional networks for semantic segmentation.
\newblock  {IEEE} Conference on Computer Vision and Pattern Recognition, {CVPR}
  2015, Boston, MA, USA, June 7-12, 2015. {IEEE} Computer Society,  2015, pp.
  3431--3440.
\newblock
  doi:{\changeurlcolor{black}\href{https://doi.org/10.1109/CVPR.2015.7298965}{\detokenize{10.1109/CVPR.2015.7298965}}}.

\bibitem[He \em{et~al.}(2015)He, Zhang, Ren, and Sun]{DBLP:conf/iccv/HeZRS15}
He, K.; Zhang, X.; Ren, S.; Sun, J.
\newblock Delving Deep into Rectifiers: Surpassing Human-Level Performance on
  ImageNet Classification.
\newblock  2015 {IEEE} International Conference on Computer Vision, {ICCV}
  2015, Santiago, Chile, December 7-13, 2015. {IEEE} Computer Society,  2015,
  pp. 1026--1034.
\newblock
  doi:{\changeurlcolor{black}\href{https://doi.org/10.1109/ICCV.2015.123}{\detokenize{10.1109/ICCV.2015.123}}}.

\bibitem[Dauphin \em{et~al.}(2017)Dauphin, Fan, Auli, and
  Grangier]{DBLP:conf/icml/DauphinFAG17}
Dauphin, Y.N.; Fan, A.; Auli, M.; Grangier, D.
\newblock Language Modeling with Gated Convolutional Networks.
\newblock  Proceedings of the 34th International Conference on Machine
  Learning, {ICML} 2017, Sydney, NSW, Australia, 6-11 August 2017; Precup, D.;
  Teh, Y.W., Eds. {PMLR},  2017, Vol.~70, {\em Proceedings of Machine Learning
  Research}, pp. 933--941.

\bibitem[Pons~Puig \em{et~al.}(2019)Pons~Puig et~al.]{pons2019deep}
Pons~Puig, J.; others.
\newblock Deep neural networks for music and audio tagging.
\newblock PhD thesis, Universitat Pompeu Fabra,  2019.

\bibitem[Kingma and Ba(2015)]{DBLP:journals/corr/KingmaB14}
Kingma, D.P.; Ba, J.
\newblock Adam: {A} Method for Stochastic Optimization.
\newblock  3rd International Conference on Learning Representations, {ICLR}
  2015, San Diego, CA, USA, May 7-9, 2015, Conference Track Proceedings;
  Bengio, Y.; LeCun, Y., Eds.,  2015.

\bibitem[Gulrajani \em{et~al.}(2017)Gulrajani, Ahmed, Arjovsky, Dumoulin, and
  Courville]{DBLP:conf/nips/GulrajaniAADC17}
Gulrajani, I.; Ahmed, F.; Arjovsky, M.; Dumoulin, V.; Courville, A.C.
\newblock Improved Training of Wasserstein GANs.
\newblock  Advances in Neural Information Processing Systems 30: Annual
  Conference on Neural Information Processing Systems 2017, December 4-9, 2017,
  Long Beach, CA, {USA}; Guyon, I.; von Luxburg, U.; Bengio, S.; Wallach, H.M.;
  Fergus, R.; Vishwanathan, S.V.N.; Garnett, R., Eds.,  2017, pp. 5767--5777.

\bibitem[Nistal \em{et~al.}(2020)Nistal, Lattner, and
  Richard]{DBLP:conf/eusipco/NistalLR20}
Nistal, J.; Lattner, S.; Richard, G.
\newblock Comparing Representations for Audio Synthesis Using Generative
  Adversarial Networks.
\newblock  28th European Signal Processing Conference, {EUSIPCO} 2020,
  Amsterdam, Netherlands, January 18-21, 2021. {IEEE},  2020, pp. 161--165.
\newblock
  doi:{\changeurlcolor{black}\href{https://doi.org/10.23919/Eusipco47968.2020.9287799}{\detokenize{10.23919/Eusipco47968.2020.9287799}}}.

\bibitem[Thiede \em{et~al.}(2000)Thiede, Treurniet, Bitto, Schmidmer, Sporer,
  Beerends, and Colomes]{thiede2000peaq}
Thiede, T.; Treurniet, W.C.; Bitto, R.; Schmidmer, C.; Sporer, T.; Beerends,
  J.G.; Colomes, C.
\newblock PEAQ-The ITU standard for objective measurement of perceived audio
  quality.
\newblock {\em Journal of the Audio Engineering Society} {\bf 2000}, {\em
  48},~3--29.

\bibitem[{}International Telecommunications
  Union–Radiocommunication~{(ITU-T)}()]{mos}
{}International Telecommunications Union–Radiocommunication~{(ITU-T)},
  R.I.T.P.

\bibitem[Trabelsi \em{et~al.}(2018)Trabelsi, Bilaniuk, Zhang, Serdyuk,
  Subramanian, Santos, Mehri, Rostamzadeh, Bengio, and Pal]{trabelsi2018deep}
Trabelsi, C.; Bilaniuk, O.; Zhang, Y.; Serdyuk, D.; Subramanian, S.; Santos,
  J.F.; Mehri, S.; Rostamzadeh, N.; Bengio, Y.; Pal, C.J.
\newblock Deep Complex Networks.
\newblock  International Conference on Learning Representations,  2018.

\bibitem[Engel \em{et~al.}(2019)Engel, Agrawal, Chen, Gulrajani, Donahue, and
  Roberts]{DBLP:conf/iclr/EngelACGDR19}
Engel, J.H.; Agrawal, K.K.; Chen, S.; Gulrajani, I.; Donahue, C.; Roberts, A.
\newblock GANSynth: Adversarial Neural Audio Synthesis.
\newblock  7th International Conference on Learning Representations, {ICLR}
  2019, New Orleans, LA, USA, May 6-9, 2019. OpenReview.net,  2019.

\end{thebibliography}
\end{document}